\documentclass[aps,prd,amsmath,amssymb,superscriptaddress,nofootinbib,onecolumn,10pt,showpacs]{revtex4}
\usepackage[]{latexsym}
\usepackage[english]{babel}
\usepackage{graphicx}
\usepackage{hyperref}
\usepackage{color}
\usepackage{bm}

\allowdisplaybreaks

\newcommand{\T}{\mathbf{\hat{\mathcal{T}}}}

\newcommand{\Dk}[1]{\frac{d^3#1}{(2\pi)^3}} 
 
\newcommand{\br}{\bar{\rho}}
\newcommand{\ve}[1]{{\text{\bf #1}}} 
\newcommand{\vk}{\ve k}
\newcommand{\vp}{\ve p}
\newcommand{\vq}{\ve q}
\newcommand{\vx}{\ve x}

\newcommand{\lcdm}{$\Lambda$CDM }

\newcommand{\obd}{\omega_\text{BD}}

\newcommand{\ikk}{\underset{\vk_{12}= \vk}{\int}}
\newcommand{\ikkk}{\underset{\vk_{123}= \vk}{\int}}

\def\typomode{0}
\usepackage{xcolor}
\definecolor{typoold}{gray}{0.55}
\definecolor{tocheck}{rgb}{0.95,0.45,0.0}
\newcommand{\typo}[2]{\ifcase\typomode\relax{#2}%
  \or{\color{red}#2}%
  \or{\color{typoold}\ensuremath{\lbrack}#1\ensuremath{\rbrack}}{\color{red}#2}\fi}
\newcommand{\tocheck}[1]{\ifcase\typomode\relax{#1}\else{\color{tocheck}#1}\fi}

\begin{document}
\title{Lagrangian perturbation theory for modified gravity}
\author{Alejandro Aviles}
\email{avilescervantes@gmail.com}
\affiliation{Departamento de F\'isica, Instituto Nacional de Investigaciones Nucleares,
Apartado Postal 18-1027, Col. Escand\'on, Ciudad de M\'exico,11801, M\'exico.}
\affiliation{Consejo Nacional de Ciencia y Tecnolog\'ia, Av. Insurgentes Sur 1582,
Colonia Cr\'edito Constructor, Del. Benito \typo{Jurez}{Ju\'arez}, 03940, Ciudad de M\'exico, M\'exico}
\author{Jorge L. Cervantes-Cota}
\email{jorge.cervantes@inin.gob.mx}
\affiliation{Departamento de F\'isica, Instituto Nacional de Investigaciones Nucleares,
Apartado Postal 18-1027, Col. Escand\'on, Ciudad de M\'exico,11801, M\'exico.}
\pacs{PACS}

\begin{abstract}
We present a formalism to compute Lagrangian displacement fields for a wide range of cosmologies in the context of 
perturbation theory up to third order. We emphasize the case of theories with scale dependent gravitational strengths, such as  
chameleons, but our formalism can be accommodated to other modified gravity theories. 
In the non-linear regime two qualitative features arise. One, as is well known, 
is that nonlinearities lead to a screening of the force mediated by the scalar field. The second 
is a consequence of the transformation of the Klein-Gordon equation from Eulerian 
to Lagrangian coordinates, producing frame-lagging terms that are important especially at large scales, 
and if not considered, the theory does not reduce to the $\Lambda$CDM model in that limit.
We apply our formalism to compute the 1-loop power spectrum and the correlation function in $f(R)$ gravity by using 
different resummation schemes. We further discuss the IR divergences of these formalisms.
\end{abstract}

\maketitle

\begin{section}{Introduction}
General Relativity (GR) is the most successful theory of gravity we have nowadays. Indeed, 
it has passed so far all experimental tests ranging from 1$\mu$m to 1 AU $\sim 10^{11}$m \cite{Will:2014kxa,Berti:2015itd}.  
Nevertheless, mainly since the discovery of the 
accelerated expansion of the universe \cite{Perlmutter:1998np,Riess:1998cb}, concerns about its validity at cosmological scales have arisen. 
In addition, with the recent and 
forthcoming  advent of a lot of cosmological data, especially 
from modern surveys \cite{Beutler:2016ixs,Garcia-Fernandez:2016oud,Aghamousa:2016zmz,Racca:2016qpi,Abell:2009aa}, 
there is an increasing interest to test gravity at cosmological scales.

The minimal modification to the Einstein-Hilbert action compatible with 
the principle of generalized covariance ---and the only that introduces no additional degrees of freedom--- is provided by the introduction of 
a cosmological constant; this leads to the standard model of Cosmology, 
the $\Lambda$CDM model, that can explain the cosmic background radiation anisotropies with exquisite precision \cite{Ade:2015xua}, 
the late time background evolution probed by measurements of the baryon acoustic oscillations (BAO)
and supernova distance-redshift relations \cite{Betoule:2014frx}, among several other observations. 
But this success comes with a price, the $\Lambda$ side of the theory suffers from the 
smallness and coincidence problems \cite{Weinberg:1988cp,Martin:2012bt}. 
One alternative to the cosmological constant is that of evolving dark energy, 
in which exotic matter fields endowed with a negative pressure counteract the gravitational 
attraction and generate the cosmic acceleration \cite{Ratra:1987rm,Tsujikawa:2010sc}. 
From the observational point of view, 
a practical inconvenience of evolving dark energy is that a background expansion of the Universe that is close to $\Lambda$CDM implies that collapse
features are nearly indistinguishable as well, as long as the dark energy is stress-free, and as a consequence it would be difficult to call the
correct theory  \cite{Bertschinger:2006aw,Bertschinger:2008zb}. 
The case of Modified Gravity (MG) is different because the two scalar gravitational 
potentials of the metric are not 
necessarily equal even in the absence of stress sources, and therefore a theory that mimics a dark energy model at the background 
level can predict a very different structure formation history. 

The construction of alternatives theories of gravity has been a difficult endeavor because they are restricted to reduce to GR in some limits, 
while still having an impact on cosmic scales in order to provide the cosmic acceleration. 
For this reason, cosmologists have concentrated in gravitational 
theories that poses a {\it screening} mechanism \cite{Khoury:2010xi}. In 
theories with one additional degree of freedom 
this can be achieved by the non-linearities of its evolution equation ---the Klein-Gordon equation in the case of a scalar field. 
The general idea is that in regions with high density or large gradients of it, 
the additional (the fifth) force becomes negligible, recovering GR. These effects can be 
achieved by the scalar field itself or by its spatial derivatives, leading to different types of screenings: 
as, for example, the chameleon \cite{Khoury:2003aq,khoury:2003rn}, kinetic \cite{Babichev:2009ee}, 
or Vainshtein \cite{Vainshtein:1972sx} mechanisms.

In this work we are interested in the formation of cosmological structures at large scales. 
In the standard scenario, this process is driven by gravitational collapse of cold dark matter (CDM). 
Once baryons decouple from the primordial plasma, they fall into potential wells already formed by the CDM, 
and ultimately forming the galaxies we observe today.
$N$-body simulations are the most trustable method to study this process, especially at small scales,
and it is considered in some sense to lead to the correct solution to the 
problem. Full $N$-body simulations have been provided for a few MG theories in \cite{Zhao:2010qy,BLi2013,Puchwein:2013lza,Llinares:2013jza,Rizzo:2016mdr,Winther:2015wla}, while semi-analytical-$N$-body codes 
that rely on screening for spherical configurations \cite{Winther:2014cia}, and the COLA approach, introduced in 
\cite{Tassev:2013pn}  for $\Lambda$CDM, were implemented recently in \cite{Valogiannis:2016ane,Winther:2017jof}.
The main disadvantage of simulations is that they are computationally very expensive; but furthermore, the understanding of how the process of structure formation
takes place stay relatively hidden. For these reasons, analytical and semi-analytical methods have been developed during the last few decades; see 
\cite{Sahni:1995rm,Ber02,Cooray:2002dia} for reviews in standard cosmology, and Refs.~\cite{Clifton:2011jh,Koyama:2015vza,Joyce:2016vqv} for MG. 

Of our particular interest in this work are perturbation 
theories of large scale structure formation, here the relevant fields are perturbed about their background cosmological values and 
evolution equations are provided by taking moments of the Boltzmann equation. 
Perturbation theory comes in different versions, each one having its advantages and disadvantages, 
mainly Lagrangian Perturbation Theory (LPT) 
\cite{Zel70,Buc89,Mou91,Catelan:1994ze,BouColHivJus95,TayHam96,Mat08a,Carlson:2012bu,Sugiyama:2013mpa,Vlah:2014nta,Matsubara:2015ipa} 
and Eulerian Standard Perturbation Theory (SPT) 
\cite{Juszkiewicz01121981,Vishniac01061983,Fry:1983cj,Goroff:1986ep,Jain:1993jh,Carlson:2009it}, 
but so many others have been tailored in order to improve
its performance, for example see Refs.~\cite{McDonald:2006hf,Crocce:2005xy,Taruya:2007xy,Pietroni:2008jx}. 
The main drawback of perturbation theory is that the expansion of fields become meaningless when higher orders become larger than lower orders, 
this occurring as soon as non-linearities become important, but at the weakly non-linear scales it leads to notable improvements 
over linear theory.

In MG, perturbation theory for large scale structure formation was developed first in \cite{2009PhRvD..79l3512K} based in the formalism of 
the closure theory of Ref.~\cite{Taruya:2007xy}. That work introduces a Fourier expansion
for the screening in which each order carries its own screening, becoming evident the effect of the screening even at large scales. 
Other works in perturbation theory for MG, some of them including the systematic effects of redshift space distortions (RSD), were presented in 
\cite{Taruya:2013quf,Brax:2013fna,Taruya:2014faa,2014PhRvD..89d3509T,Bellini:2015oua,Taruya:2016jdt,Bose:2016qun,Fasiello:2017bot,Bose:2017dtl}. 
The formalisms developed so far are closer to SPT than to LPT. 
An advantage of LPT is that it is better in modeling the BAO features of the power spectrum, and 
it is highly successful in reproducing the correlation function around the acoustic peak \cite{Tassev:2013rta}. 
Formally, the power spectrum derived from SPT cannot 
be Fourier transformed to obtain the correlation function, but even if a damping factor is introduced to make the integral convergent a double peaked structure 
is observed around the BAO scale. On the other hand, SPT follows better than LPT the broadband power spectrum trend line; in the latter 
it decays very quickly at small scales because dark matter particles follow their trajectories almost inertially and do not clump enough to form small
structures leading to a deficit in the power spectrum \cite{Vlah:2014nta}.

In this work we develop an LPT theory that can be applied to a wide range of MG theories, essentially to those that can be written as a 
scalar tensor gravity theory. When the Klein-Gordon equation for the scalar field is expressed in Lagrangian coordinates additional terms
arise that can not be neglected, otherwise GR is not recovered at large scales, we treat in detail these contributions.  
Throughout this work we exemplify the formalism by applying it to the Hu-\typo{Sawicky}{Sawicki} $f(R)$ model \cite{Hu:2007nk}.
Our method can also be applied straightforwardly to dark energy models in which we can neglect the
dark energy perturbations; and, by adding a source to the Lagrangian displacement equations these perturbations can be incorporated
if they remain linear. Though, we do not work out these ideas here. 

Since we are interested in MG models that preserve the equivalence
principle, at least at the large scales, matter and Lagrangian displacements statistics 
should be IR-safe as it is know from the works of 
Refs.~\cite{1996ApJ...456...43J,Scoccimarro:1995if,Peloso:2013zw,Carrasco:2013sva,Creminelli:2013mca}. 
Therefore, we further discuss potentially undesirable divergences during this work.

Two different second order LPT theories (2LPT) for modified gravity has been recently presented in \cite{Valogiannis:2016ane,Winther:2017jof}
in order to provide a framework for dealing with large scales in the COLA code. More recently, the 2LPT of \cite{Winther:2017jof} was extended
to include massive neutrinos in \cite{Wright:2017dkw}.  Nevertheless the formalisms of these works differ. We expect our results can clarify these discrepancies.

The rest of the paper is organized as follows: In Sect.\ref{Sect:STG} we give a brief introduction to MG theories; 
in Sect.\ref{Sect:gentheory} we present the general LPT theory; in Sect.\ref{Sect:2LPT} we show the results for the second order
2LPT theory; Sect.\ref{Sect:3LPT} is devoted to the third order; in Sect.~\ref{Sect:LDstats} we study Lagrangian displacement statistics relevant 
to obtain the matter power spectrum, and discuss their IR and UV divergences; in Sect.\ref{Sect:2pStats} we compute the power spectra and correlation
function using different resummation schemes and further discuss their properties; 
and in Sect.\ref{Sec:concl} we present our conclusions. Some computations are delegated to two appendices.

\end{section}

\begin{section}{Theories of modified gravity}\label{Sect:STG}

The schemes treated in this work are general modified gravity theories that \emph{i)} have a screening mechanism to comply with local, 
gravitational constraints; \emph{ii)} have a fifth force due to a scalar degree of freedom that acts at intermediate scales; 
and \emph{iii)} at large, cosmological 
distances GR is recovered. General modified gravity scenarios and their cosmological consequences have been reviewed in recent 
years  \cite{Clifton:2011jh,Joyce:2016vqv,Koyama:2015vza}, from which we extract the models we are interested in. 

We may consider general scalar-tensor theories \cite{Bergmann:1968ve,1970ApJ...161.1059N,PhysRevD.1.3209}, 
\begin{equation}
\label{ST1}
{\cal L}_{ST} = \frac{1}{16 \pi} \sqrt{-g} \left[ f(\varphi) R-g(\varphi) \nabla_\mu \varphi \nabla^{\mu}\varphi - V(\varphi) \right] +
{\cal L}_m (\Psi, h(\varphi) g_{\mu\nu}),
\end{equation}
where $f, g, h,$ and $V$ are functions that can be different for various theoretical frameworks.  
A conformal transformation $h(\varphi) g_{\mu\nu} \rightarrow g_{\mu\nu}$ decouples matter from the scalar field, and therefore in this {\it Jordan Frame}
point particles follow geodesics of the new metric.  One can further redefine $f, g, h,$ and $V$ to have a non-minimal coupling given by

\begin{equation}\label{ST2}
\mathcal{L} = \frac{1}{16 \pi G} \sqrt{-g} \left[\varphi R-\frac{\obd(\varphi)}{\varphi} \nabla_\mu \varphi \nabla^{\mu}\varphi -V (\varphi)\right] +
\mathcal{L}_m(\Psi, g_{\mu\nu}),
\end{equation}
which reproduces the Brans-Dicke theory when $\obd$ is a constant and $V = 0$.  It is known that the Lagrangian density of Eq. (\ref{ST2}) can be 
transformed to the {\it Einstein Frame} (in which the Ricci scalar is not coupled to scalar fields) by means of yet another 
conformal transformation. 
In this way, it is common 
to study cosmological physics in the Einstein Frame.

\bigskip 

Another common type of theory is the $f(R)$ gravity \cite{1970MNRAS.150....1B,1980PhLB...91...99S}, 
in which one replaces the Einstein-Hilbert Lagrangian density by a generic function of the Ricci scalar, such that
\begin{equation}
\mathcal{L}_{R} =  \sqrt{-g} (R + f(R)) .  \label{f_of_R}
\end{equation}
This Lagrangian can also be transformed to different frames by using two different formalisms: a Legendre transformation relates Eq. (\ref{f_of_R}) 
to a scalar tensor theory with null coupling parameter $\obd$ and with no couplings to possibly added matter fields in Eq. (\ref{f_of_R}). The other way is to 
again apply a conformal transformation to obtain a scalar tensor theory in the Einstein frame \cite{BARROW1988515,PhysRevD.39.3159}.

\bigskip 

The Lagrangian densities mentioned above are the most considered in the literature. Thus, when studying screening mechanisms for the scalar field, one 
can consider the Einstein frame and scalar field Lagrangian of the following general form \cite{Joyce:2014kja,Koyama:2015vza},
\begin{equation}
 {\cal L} = -\frac{1}{2}Z^{\mu\nu}(\varphi, \partial\varphi,\partial^2 \varphi)\partial_\mu\varphi\partial_\nu\varphi-V(\varphi)+ \beta(\varphi)T^\mu_{\;\mu} ,
\label{screen_lag}
\end{equation}
where the function $Z^{\mu\nu}$ stands for possible derivative self-interactions of the scalar field, $V(\varphi)$ its 
a potential, and $T^\mu_{\;\mu}$ is the trace of the 
matter stress-energy tensor.  The coupling $\beta$ yields a scalar field solution that depends on the local density; for non-relativistic matter 
$T^\mu_{\;\mu}= - \rho$.   Eq. (\ref{screen_lag}) is then a generalization to Eq. (\ref{ST2}) in which one introduces more general scalar field 
kinetic terms. Equation (\ref{screen_lag}) encompasses the most used modified gravity theories ---including Horndeski theories \cite{Horndeski:1974wa}--- 
satisfying known observational constraints but allowing 
smoking guns to be discriminated or confirmed by near-future cosmological data, e.g. \cite{Aghamousa:2016zmz}.  

The screening of the fifth force at local 
scales can be realized when considering fluctuations around background values with mass $m(\varphi)$ and due to the presence of the functions 
$Z^{\mu\nu}$ or $\beta$. According to Ref. \cite{Joyce:2014kja}, the screening can be classified as due to: i) {\it Weak coupling}, in which the function $\beta$ is
small in regions of high density (local scales), thus the fifth force is suppressed. At large scales the density is small and the fifth forces acts. Examples 
of theories of this type are symmetron \cite{PhysRevLett.104.231301,PhysRevD.72.043535,PhysRevD.77.043524} and varying 
dilaton \cite{DAMOUR1994532,PhysRevD.83.104026}; ii)  {\it Large mass}, when mass of the fluctuation is large in regions of high density and the fifth force 
is suppressed, and at low densities the scalar field is small and can mediate a fifth force.  Examples of this type are 
Chameleons \cite{Khoury:2003aq,khoury:2003rn}; and iii) {\it Large inertia}, when the kinetic function $Z$ depends on the environment, making it large 
in density regions, and then the coupling to matter is suppressed. One finds two cases, when the first derivatives of the scalar field are large or when the second
derivatives are large. Examples of the former are $k$-mouflage models \cite{Babichev:2009ee,PhysRevD.84.061502}, and of the latter are 
Vainshtein models \cite{Vainshtein:1972sx}. 

\bigskip

Let us now focus on perturbations. When considering quasi-nonlinear scales, it is common to use the Jordan frame, as given by 
Eq.~(\ref{ST2}), following \cite{2009PhRvD..79l3512K}. Since our interest will be the study of kinematics well inside the cosmological 
horizon, it seems plausible to take the quasi-static limit, in which time derivatives of the scalar field are much less important than 
spatial derivatives. This approximation is non-trivial however, as it has been shown in the context of modified gravity models  
\cite{Brax:2013fna,PhysRevD.92.084061,Noller:2013wca,Bose:2014zba}, but its correctness depends upon details 
of the modified gravity free parameters which vary from model to model. Therefore, it is improper to assume its correctness in general, and in fact its accuracy 
depends upon model parameters and wavenumbers studied. For the concrete models we will later treat in this work (Hu-\typo{Sawicky}{Sawicki} gravity) the validity of  
this approximation has been checked within the linear perturbation theory in  \cite{Noller:2013wca} and numerical simulations in \cite{Bose:2014zba}, 
and in a general framework, introducing the effective sound speed ($c_s$) of the $f(R)$ gravity model in \cite{Brax:2013fna};  this later 
work concludes that the quasi-static approximation depends, not on the wavenumber in units of the cosmological horizon, $k/aH$, but on the combination $c_s k/aH$, 
which must be greater than unity.\footnote{In realistic, observational applications this depends on the volume spanned by surveys, 
and the redshifts to be considered. The bigger and deeper 
the survey the closer to unity the sound speed must be.}  Assuming these considerations are taken into account, we consider perturbations in a standard way 
using the Newtonian gauge, one end up with the following set of MG equations \cite{2009PhRvD..79l3512K}: 
\begin{align}
\frac{1}{a^2} \nabla^2 \psi &= 4 \pi G \bar{\rho} \delta -\frac{1}{2 a^2} \nabla^2 \varphi,\label{grav2} \\
(3 +2 \obd) \frac{1}{a^2} \nabla^2 \varphi &= - 8 \pi G \bar{\rho} \delta + {\rm NL},
\end{align}
where $\bar{\rho}$ is the background matter energy density, $\delta$ its perturbation and $\psi$ the gravitational potential. NL stands for possible 
non-linearities that may appear in the Lagrangian.  As usual, these equations are coupled to the standard continuity and Euler 
equations for the evolution of the matter fields. 

When going to the Fourier space, the Klein-Gordon  equation can be written as
\begin{equation}
(3 +2 \obd) \frac{k^2}{a^2} \varphi
= 8 \pi G \typo{\rho}{\bar{\rho}} \delta - {\cal I}(\varphi). \label{KG_eq}
\end{equation}
Here the self-interaction term ${\cal I}$ may be expanded following \cite{2009PhRvD..79l3512K} as 
${\cal I} (\varphi) = M_1(k) \varphi  + \delta \mathcal{I}(\varphi)$ with
\begin{align}\label{selfIntexp}
\delta \mathcal{I}(\varphi)
&= \frac{1}{2} \int \frac{d^3 k_1 d^3 k_2}
{(2 \pi)^3} \delta_D(\vk -\vk_{12}) M_2(\vk_1, \vk_2)
\varphi(\vk_1) \varphi(\vk_2) \nonumber\\
&\quad + \frac{1}{6}
\int \frac{d^3 k_1 d^3 k_2 d^3 k_3}{(2 \pi)^6}
\delta_D(\vk - \vk_{123}) M_3(\vk_1, \vk_2, \vk_3)
\varphi(\vk_1)\varphi(\vk_2) \varphi(\vk_3) + \cdots,
\end{align}
where the functions $M_i$  are in general scale and time dependent and are determined by the particular MG model. In the following sections 
we will keep this dependence explicitly, although we 
illustrate our results with $f(R)$ theories for which the $M_i$ are only time dependent, as we see below.  

\bigskip

Now we consider in more detail $f(R)$ gravity, that was first used
to explain the current cosmic acceleration  in \cite{Capozziello:2002rd,Carroll:2003wy}.
Variations with respect to the metric of the action constructed with the Lagrangian density of Eq.~(\ref{f_of_R}), lead to the field equations
\begin{equation}\label{fRFE}
 G_{\mu\nu} + f_R R_{\mu\nu} - \nabla_\mu \nabla_\nu f_R - \left( \frac{f}{2} - \square f_R \right)g_{\mu\nu} = 8 \pi G T_{\mu\nu},   
\end{equation}
where $f_R \equiv \frac{df(R)}{dR}$.
By taking the trace to this equation we obtain
\begin{equation} \label{TeqfR}
3 \square f_R = R(1-f_R) + 2 f - 8 \pi G \rho,
\end{equation}
where we use a dust-like fluid with $T^{\mu}_\mu = -\rho$. In cosmological \typo{pertubative}{perturbative} treatments one splits
$f_R= \bar{f_R} + \delta f_{R}$ and $R= \bar{R} + \delta R$, where the bar indicates background quantities and $\bar{R} \equiv R(\bar{f}_R)$.
Necessary conditions to get a background cosmology consistent 
with the $\Lambda$CDM model are $|\bar{f}/ \bar{R}| \ll 1$ and $|\bar{f}_R| \ll 1$ \cite{Hu:2007nk}. 

In the quasi-static limit, the trace equation can be written as
\begin{equation}\label{TeqfRql}
 \frac{3}{a^2} \nabla^2 \delta f_R = - 8 \pi G \bar{\rho}\delta + \delta R,
\end{equation}
where we \typo{substracted}{subtracted} the background terms in Eq.(\ref{TeqfR}). We recognize Eq.(\ref{TeqfRql}) as the Klein Gordon equation for a scalar field 
with a potential $\delta R$ and Brans-Dicke parameter $\obd=0$. Since $\delta R = R(f_R) - R(\bar{f}_{R})$, we may expand the potential as
\begin{equation} \label{potexp}
 \delta R = \sum_{\typo{i}{n}} \frac{1}{n!} M_n (\delta f_R)^n, \qquad M_n \equiv \frac{d^n R(f_R)}{d f_{R}^n} \Bigg|_{f_R = \bar{f}_R},
\end{equation}
from which the mass associated to the perturbed field $\delta f_R$ can be read as
\begin{equation}\label{mass}
 m = \sqrt{\frac{M_1}{3}}.
\end{equation}

In this work we focus our attention to the  Hu-\typo{Sawicky}{Sawicki} model \cite{Hu:2007nk}, defined by
\begin{equation}\label{HSfR}
 f(R)= -M^2 \frac{c_1 (R/M^2)^n}{c_2(R/M^2)^n + 1},
\end{equation}
where the energy scale is chosen to be $M^2= H_0^2 \Omega_{m0}$. In this parametrized model, at high curvature ($R\gg M^2$)
the function $f(R)$ approaches a constant, recovering GR with cosmological constant, while at low curvature it goes to zero,
recovering GR; the manner these two behaviors are interpolated is dictated by the free parameters. 
Given that $f_{RR} > 0$ for $R>M^2$ the solutions are stable 
and the mapping to a scalar tensor gravity is allowed \cite{Magnano:1993bd}. 
The latter also implies that this gravitational model introduces a single one extra degree of freedom. In order to mimic the background evolution 
of the $\Lambda$CDM model it is also necessary that $c_1/c_2 = 6 \Omega_\Lambda / \Omega_{m0}$ \cite{Hu:2007nk}, thus leaving two parameters to
fix the model.  Noting that the background value is about $\bar{R} \gtrsim
40 M^2 \gg M^2$, Eq.(\ref{HSfR}) simplifies and leads to
\begin{equation} \label{HSfrfR}
 f_R \simeq f_{R0} \left(\frac{\bar{R}}{R}\right)^{n+1},
\end{equation}
with 
\begin{equation} \label{BackRicci}
 \bar{R} = 6(\dot{H} + 2H^2) =  3 H^2_0 ( \Omega_{m0} a^{-3} + 4 \Omega_\Lambda).
\end{equation}
We consider the case $n=1$. Using Eqs.(\ref{potexp}) and (\ref{HSfrfR}) we find the functions $M_1$, $M_2$ and $M_3$:
\begin{align}
M_1(a) = \frac{3}{2}  \frac{H_0^2}{|f_{R0}|} \frac{(\Omega_{m0} a^{-3} + 4 \Omega_\Lambda)^3}{(\Omega_{m0}  + 4 \Omega_\Lambda)^2}, \label{M1}
\end{align}
\begin{align}
M_2(a) = \frac{9}{4}  \frac{H_0^2}{|f_{R0}|^2} \frac{(\Omega_{m0} a^{-3} + 4 \Omega_\Lambda)^5}{(\Omega_{m0}  + 4 \Omega_\Lambda)^4}, 
\end{align}
\begin{align}
M_3(a)= \frac{45}{8}  \frac{H_0^2}{|f_{R0}|^3} \frac{(\Omega_{m0} a^{-3} + 4 \Omega_\Lambda)^7}{(\Omega_{m0}  + 4 \Omega_\Lambda)^6}. 
\end{align}
These functions depend only on the background evolution since they are the coefficients of the expansion of a scalar field potential 
about its background value. In other theories, {\it e.g.} with non-canonical kinetic terms as \typo{Galilieons}{Galileons} or DGP, the ``potential'' includes spatial
derivatives, thus the $M_i$ functions may carry scale dependences; 
see, for example, \cite{2009PhRvD..79l3512K} for the  $M_i$ functions in DGP gravity. 


With $n$ fixed to unity, the only free parameter is $f_{R0}$. The models constructed with it are named F4, F5, F6 and so on, corresponding
to the choices $f_{R0} = -10^{-4}, -10^{-5}, -10^{-6}$ respectively. From these, F4 represents the largest deviation to GR, and 
though it is ruled out using linear theory, it is worthy to analyze its effects. 
These theories are indistinguishable from GR at the background, homogeneous and 
isotropic cosmic evolution \cite{Hu:2007nk} as already was assumed in Eq.~(\ref{BackRicci}). 
Therefore throughout this work, we shall use these three models to exemplify the 
method with an expansion history given by a $\Lambda$CDM model with $\Omega_m=0.281$ and $h = 0.697$.

In this work we specialize to $f(R)$ theories. Even though, by dealing consistently with the scale dependence of the $M_i$ 
functions, our results cover a wide range of models, essentially the whole Horndeski sector. Indeed, in appendix B of Ref.~\cite{Bose:2016qun} it is
shown how to relate the $M_i$ functions to the Horndeski free functions.\footnote{More precisely, 
in that work the authors use the so-called $\gamma_1$ and $\gamma_2$ kernels
for screenings, these are obtained by expanding  $\delta I$  in terms of overdensities, 
instead of scalar fields as in Eq.~(\ref{selfIntexp}). That is, the $\gamma_i$
functions are the kernels of the expansion of Eq.~(\ref{dIexp}) below.}

\end{section}

\begin{section}{Lagrangian perturbation theory in modified gravity}\label{Sect:gentheory}

In a Lagrangian description we follow the trajectories of individual particles with initial position $\vq$ 
and current position $\vx$. Lagrangian and Eulerian coordinates are related by the transformation rule
\begin{equation} \label{LagtoEul}
 \vx(\vq,t) = \vq + \mathbf{\mathcal{S}}(\vq,t),
\end{equation}
with the Lagrangian map $\mathbf{\mathcal{S}}(\vq,t)$ vanishing at the initial time where 
Lagrangian coordinates are defined. In kinetic theory we define the Lagrangian displacement
velocity $\dot{\Psi}(\vq,t)$ as a momentum average of particles' velocities over the ensemble.  
For CDM scenarios, which we posit here, we can safely neglect velocity dispersions and
the identity $\mathcal{S} = \mathbf{\Psi}$ follows  \cite{Aviles:2015osc}. Furthermore,
mass conservation implies the relation between Lagrangian displacements and overdensities
\begin{equation} \label{reldJ}
\delta(\vx,t) = \frac{1-J(\vq,t)}{J(\vq,t)},
\end{equation}
where the Jacobian of the transformation is
\begin{equation}\label{jac}
 J_{ij} := \frac{\partial x^i}{\partial q^j} = \delta_{ij} + \frac{\partial \Psi^i}{\partial q^j}
\end{equation}
and $J$ is its determinant. We make note that in writing Eq.~(\ref{reldJ}) we identify the Lagrangian displacement field
as a random field. We further assume this field to be Gaussian at first order in perturbation theory.

We consider the cases for which the MG theory can be written as a Brans-Dicke like model at linear level.   
For longitudinal modes we have to solve the system
\begin{equation} \label{longLPTeq}
 \nabla_{\vx} \cdot (\ddot{\Psi}(\vq,t) + 2 H \dot{\Psi}) = -\frac{1}{a^2}\nabla^2_\vx \psi(\vx,t)
\end{equation}
\begin{equation} \label{PoissonEq}
 \frac{1}{a^2} \nabla^2_\vx \psi = 4 \pi G \br \delta(\vx,t) -  \frac{1}{2a^2} \nabla^2_\vx \varphi.
\end{equation}
In $x$-Fourier space, and in the quasi-static limit, the Klein Gordon equation is
\begin{equation} \label{EulerianBDeq}
(3 + 2 w_\text{BD}) \frac{k_x^2}{a^2} \varphi(\vk_x,t) = 8 \pi G \br \delta(\vk_x,t) - M_1(\vk_x,t) \varphi(\vk_x,t) - \delta I (\varphi).
\end{equation}
The notation $\vk_x$ emphasizes that the wavenumbers correspond to Eulerian coordinates, 
that are not the correct coordinates to be transformed to the Fourier space, since equations of motion will be given in the $q$-space, therefore
we shall work in $q$-Fourier space,
where wavenumbers correspond to Lagrangian coordinates. For the sake of brevity, in the following we will omit the time argument $t$ when this
not lead to confusions. The term $I(\varphi) = M_1\varphi + \delta I (\varphi)$, introduced in \cite{2009PhRvD..79l3512K}, 
is the expansion of the scalar field potential about its background
value $\varphi$, being $M_1 \varphi$ the linear piece and $\delta I (\varphi)$ the non-linear which should 
give rise to the screening of the fifth force. If the latter is not present, the
solutions lead to large violations of local experiments, unless
the Brans-Dicke parameter $\obd$ is very large. On the other hand, theories which \typo{poses}{possess} a screening mechanism
shield the scalar field in nonlinear regions such that $\nabla^2_\vx \varphi \approx 0$, recovering GR. 
Following \cite{Matsubara:2015ipa}, we define the operator
\begin{equation}\label{defT}
 \T = \frac{d^2 \,}{dt^2} + 2 H \frac{d \,}{dt},
\end{equation}
thus, Eqs.~(\ref{longLPTeq}) and (\ref{PoissonEq}) can be written as 
\begin{equation}\label{longLPTeq2}
 \nabla_{\vx} \cdot \T \Psi =  -4 \pi G \br \delta(\vx,t) +  \frac{1}{2a^2} \nabla^2 \varphi 
 + \frac{1}{2a^2} \big( \nabla^2_\vx \varphi - \nabla^2 \varphi \big),
\end{equation}
where $\nabla_i = {}_{,i}$ is partial derivative with respect to $q^i$.
With this splitting of the Laplacian we recognize the scalar field as a function of Lagrangian coordinates, the term  
$\nabla^2_\vx \varphi - \nabla^2 \varphi $ has a geometrical nature that arises  when transforming the Klein-Gordon equation from
Eulerian to Lagrangian coordinates. We shall call it the {\it frame-lagging term}, and we show below that is non-negligible and it is
necessary for recovering  $\Lambda$CDM at sufficiently large scales, where the fifth force mediated by the scalar field is essentially zero.  

To transform to Lagrangian coordinates we use the relations
\begin{align}
J &= \frac{1}{6} \epsilon_{ijk} \epsilon_{pqr} J_{ip}J_{jq}J_{kr}, \label{detJ}\\
(J^{-1})_{ij} &= \frac{1}{2J} \epsilon_{jkp} \epsilon_{iqr} J_{kq}J_{pr}, \label{InvJ}
\end{align}
with $ \epsilon_{ijk}$ the fully antisymmetric Levi-Civita symbol.
In perturbation theory the relevant fields are formally expanded as
\begin{align}
 \Psi &= \lambda \Psi^{(1)} + \lambda^2 \Psi^{(2)} + \lambda^3 \Psi^{(3)}  + \mathcal{O}(\lambda^4), 
\end{align}
and analogously for $\delta$ and $\varphi$.
From now on the control parameter $\lambda$ is absorbed in the definitions of the perturbed fields. Since spatial derivatives
in Lagrangian and Eulerian coordinates are related by $\nabla_{\vx \, i} = \typo{(J^{-1})_{im}}{(J^{-1})_{mi}}\nabla_m$, 
the frame-lagging term of Eq.~(\ref{longLPTeq2}) can be written as
\begin{equation}\label{lagFrame}
 \nabla^2_{\vx} \varphi - \nabla^2 \varphi   = \typo{(J^{-1})_{im}}{(J^{-1})_{mi}}\nabla_m (\typo{(J^{-1})_{in}}{(J^{-1})_{ni}}  \nabla_n \varphi) - \nabla^2 \varphi \sim \mathcal{O}(\lambda^2),
\end{equation}
implying it is a non-linear term. The Fourier transform of $\varphi(\vq)$ leads to 
\begin{equation} \label{sfFourier}
- \frac{k^2}{2 a^2} \varphi(\vk) = - (A(k)-A_0) \tilde{\delta}(\vk) + \frac{k^2/a^2}{6 \Pi(k)} \delta I (\vk)
- \frac{(3 + 2 \obd) k^2/a^2}{3 \Pi(k)}  \frac{1}{2a^2} [(\nabla^2_\vx \varphi - \nabla^2 \varphi)](\vk),
\end{equation}
where $[\,(\,\cdots\,)\,](\vk)$ means Fourier transform of $(\,\cdots\,)(\vq)$ and we defined
\begin{align}
 A(k) &= 4 \pi G \bar{\rho}  \left( 1 + \frac{k^2/a^2}{3 \Pi(k)} \right) \label{defA}, \\
 \Pi(k) &= \frac{1}{3 a^2}\big((3 + 2 \obd) k^2 + M_1 a^2 \big) \label{defPi}, \\
 A_0 &= A(k=0,t) =  4 \pi G \bar{\rho}. \label{defA0}
\end{align}
$A(k)$ is the gravitational strength in the MG cases, while $A_0$ is for GR.

\bigskip

In general we do not write
a tilde over Fourier space functions, since it does not lead to confusion, but we do write a tilde 
over the overdensity $\delta$ 
to make clear that the Fourier transform of $\delta(\vx)$ is taken with Lagrangian coordinates: it is neither the $q$-Fourier transform of 
$\delta(\vq)$ nor the $x$-Fourier transform of 
$\delta(\vx)$. It is given instead by $\tilde{\delta}(\vk) = \int d^3 q e^{-i\vk \cdot \vq} \delta(\vx)$, or using Eq.~(\ref{reldJ}) this is
\begin{equation}
 \tilde{\delta}(\vk) = \left[ \frac{1-J(\vq)}{J(\vq)} \right](\vk).
\end{equation}
Only for the linear overdensities we have $\tilde{\delta}^{(1)}(\vk) = \delta^{(1)}(\vk_x) = \int d^3 x e^{-i\vk \cdot \vx} \delta^{(1)}(\vx)$.

The function $M_1$ is related to the mass
of the scalar field by $M_1 = (3 +2 \obd) m^2(a)$, which gives the range of the interaction $1/m(a)$, if it is finite we recover GR at large scales. 
At sufficiently small scales, such that $M_1 \ll k^2/a^2$, we note 
$A(k) \rightarrow (4 + 2 \obd)/(3 + 2 \obd) A_0$, for which solar system observations restrict $\obd > 40000$ \cite{Will:2014kxa}, 
making the theory effectively indistinguishable from GR in the absence of non-linear terms. For example,
in $f(R)$ gravity the scalar field perturbation is identified with $\delta f_R$ as it is discussed in Sec.\typo{\ref{Sect:gentheory}}{\ref{Sect:STG}}, obtaining
$\obd =0$ for the Brans-Dicke parameter as can be seen 
directly from Eq.(\ref{TeqfRql}). Therefore, in this situation  $A(k\rightarrow \infty) \rightarrow \frac{4}{3} A_0$ which would rule out the theory; nevertheless,
the nonlinearities of the potential (encoded in $\delta I$ in perturbation theory) 
are responsible to drive the theory to GR in that limit \cite{Hu:2007nk}. On the other hand, the large scale behavior is dictated by the mass $m$,
related to $M_1$ in Eq.~(\ref{mass}),  and if it is not 
zero, GR is recovered at large scales.

Now, using Eqs.(\ref{jac}), (\ref{longLPTeq2}), and (\ref{sfFourier}), we obtain the equation of motion
\begin{equation} \label{preEqM}
 \big[ \typo{(J^{-1})_{ij}}{(J^{-1})_{ji}} \T \Psi_{i,j} \big](\vk) = - A(k) \tilde{\delta}(\vk) 
 + \frac{k^2/a^2}{\typo{3}{6} \Pi(k)} \delta I (\vk)  +  \frac{M_1}{3 \Pi(k)}   \frac{1}{2a^2} [(\nabla^2_\vx \varphi - \nabla^2 \varphi)](\vk),
\end{equation}
and up to third order the frame-lagging term is 
\begin{equation} \label{FLexp}
 [(\nabla^2_\vx \varphi - \nabla^2 \varphi)](\vk) = [- 2\Psi_{i,j}\varphi_{,ij} - \Psi_{i,ij}\varphi_{,j} +
 3  \Psi_{i,j}\Psi_{j,k}\varphi_{,ki} + 2 \Psi_{i,j}\Psi_{j,ik}\varphi_{,k} + \Psi_{l,l i}\Psi_{i,j}\varphi_{,j}](\vk), 
\end{equation}
which is obtained from Eqs.~(\ref{InvJ}) and (\ref{lagFrame}). We see below that this term has a key role to 
understand the correct contributions to LPT. We write it as a Fourier expansion as 
follows\footnote{Throughout this paper we adopt the shorthand notations  
\begin{equation}
 \underset{\vk_{12\dots n}= \vk}{\int}  = \int \prod_{i=1}^{n} \frac{d^3 k_i}{(2\pi)^3} (2\pi)^3 \delta_D(\ve k - \vk_{12\cdots n}).  
\end{equation}
and $\vk_{12\dots n}= \vk_1 + \vk_2 +\cdots \vk_n$.}  
\begin{align}
\frac{1}{2a^2}[(\nabla^2_\vx \varphi - \nabla^2 \varphi)](\vk) &= 
-\frac{1}{2}  \ikk \mathcal{K}^{(2)}_\text{FL}(\vk_1,\vk_2)  \delta^{(1)}(\vk_1,t) \delta^{(1)}(\vk_2,t)  \nonumber\\*
& \,\, - \frac{1}{6} \ikkk\mathcal{K}^{(3)}_\text{FL}(\vk_1,\vk_2,\vk_3) \delta^{(1)}(\vk_1,t) \delta^{(1)}(\vk_2,t) \delta^{(1)}(\vk_3,t).
\end{align} 
Below we give expressions for kernels $\mathcal{K}_\text{FL}$. By using Eq.~(\ref{sfFourier}) iteratively order by order 
we can write Eq.(\ref{selfIntexp}) as
\begin{align} \label{dIexp}
 \delta I (\vk) 
 &= \frac{1}{2} \left( \frac{2 A_0}{3} \right)^2 \ikk M_2(\vk_1,\vk_2) \frac{\tilde{\delta}(\vk_1) \tilde{\delta}(\vk_2)}{\Pi(k_1) \Pi(k_2)} 
\,+ \, \frac{1}{6} \left( \frac{2 A_0}{3} \right)^3 \ikkk \Bigg( M_3(\vk_1,\vk_2,\vk_3)   \nonumber\\ 
 & \qquad \qquad- \frac{ M_2(\vk_1,\vk_{23})(M_2(\vk_2,\vk_3)+J^{(2)}_{FL}(\vk_2,\vk_3)(3+ 2 w_\text{BD}))}{\Pi(k_{23})} \Bigg)
 \frac{\tilde{\delta}(\vk_1) \tilde{\delta}(\vk_2)\tilde{\delta}(\vk_3)}{\Pi(k_1) \Pi(k_2)\Pi(k_3)} + \cdots. 
\end{align}
For compactness we introduced the function
\begin{equation}
 J_{FL}^{(2)}(\vk,\vp) = 2 \left(\frac{3}{2A_0}\right)^{2}\mathcal{K}^{(2)}_\text{FL}(\vk,\vp) \Pi(k) \Pi(p).
\end{equation}
In LPT it is usual to multiply the equation of motion (Eq.~(\ref{longLPTeq2})) by the determinant $J$ before performing 
the Fourier transform, leading to a closed equation of motion cubic in $\mathbf{\Psi}$  \cite{Matsubara:2015ipa}. 
Since in our case the gravitational strength is scale-dependent, that approach leads to further complications. We instead 
use Eq.~(\ref{preEqM}) and expand quantities as 
\begin{equation}\label{expJinv}
 (J^{-1})_{ij} = \delta_{ij} - \Psi_{i,j} + \Psi_{i,k}\Psi_{k,j} + \mathcal{O}(\lambda^3),
\end{equation}
\begin{align}
 J = 1 + J_1 + J_2 + J_3, \quad
  \frac{J-1}{J} = J_1 + J_2 - J_1^2 + J_3 - 2 J_1J_2 + J_1^3 + \mathcal{O}(\lambda^4),
\end{align}
where
\begin{align}
 J_1 = \Psi_{i,i}, \quad J_2 = \frac{1}{2}( (\Psi_{i,i})^2 - \Psi_{i,j}\Psi_{j,i} ), \quad
 J_3 = \frac{1}{6}(\Psi_{i,i})^3 -  \frac{1}{2}\Psi_{i,i} \Psi_{j,k}\Psi_{k,j} + \frac{1}{3}\Psi_{i,j} \Psi_{j,k}\Psi_{k,i}.
\end{align}
We are interested in finding an equation valid up to third order because their solutions lead to the first corrections to the 
power spectrum (the 1-loop) and correlation function. The inverse of the Jacobian matrix was expanded up to second order because it is
already multiplied by the Lagrangian displacement in Eq.~(\ref{preEqM}). Now, using Eq.~(\ref{reldJ}), the matter perturbation is
\begin{align} \label{deltax}
 -\delta(\vx) &= \frac{J(\vq)-1}{J(\vq)} = \Psi_{i,i} -\frac{1}{2}( (\Psi_{i,i})^2 + \Psi_{i,j}\Psi_{j,i} ) 
 + \frac{1}{6}(\Psi_{i,i})^3  + \frac{1}{3}\Psi_{i,j} \Psi_{j,k}\Psi_{k,i} + \frac{1}{2}\Psi_{i,i} \Psi_{j,k}\Psi_{k,j} +  \mathcal{O}(\lambda^4).
\end{align}
Noting from Eq.~(\ref{expJinv}) that $\typo{(J^{-1})_{ij}}{(J^{-1})_{ji}} \T \Psi_{i,j} = \T \Psi_{i,i} - \Psi_{i,j}\T \typo{\Psi_{i,j}}{\Psi_{j,i}} + 
 \Psi_{i,k}\Psi_{k,j} \T \typo{\Psi_{i,j}}{\Psi_{j,i}}$, and using Eqs.~(\ref{preEqM}) and (\ref{deltax}) we find the Lagrangian displacement equation 
 in Fourier space for third order perturbation theory:
\begin{align} \label{eqm}
 (\T - A(k)) [\Psi_{i,i}](\vk) &= [\Psi_{i,j} \T \Psi_{j,i}](\vk) 
 - \frac{A(k)}{2}[\Psi_{i,j} \Psi_{j,i}](\vk) - \frac{A(k)}{2} [(\Psi_{l,l})^2](\vk)  \nonumber\\*
 & - [\Psi_{i,k}\Psi_{k,j} \T \Psi_{j,i}](\vk) +\frac{A(k)}{6} [(\Psi_{l,l})^3](\vk) 
  + \frac{A(k)}{2} [\Psi_{l,l} \Psi_{i,j} \Psi_{j,i}](\vk) \nonumber\\*
  &+ \frac{A(k)}{3} [\Psi_{i,k}\Psi_{k,j} \Psi_{j,i}](\vk) 
  + \frac{k^2/a^2}{6 \Pi(\vk)} \delta I (\vk)  + \frac{M_1}{6 \Pi(k)} \frac{1}{a^2} [(\nabla^2_\vx \varphi - \nabla^2 \varphi)](\vk).
\end{align}

\bigskip

At linear order the right hand side of Eq.~(\ref{eqm}) is set to zero, leading to the Zel'dovich solution
\begin{equation}\label{1LD}
 \Psi^{i}(\vk,t) = i \frac{k^{i}}{k^2} D_+(k,t) \delta^{(1)}(\vk,t=t_0),
\end{equation}
where we choose $t_0$ to be the present time and $D_+(k)$ is the fastest growing solution to 
\begin{equation} \label{LinearLPTeq}
 (\T - A(k))D_+(k) = 0,
\end{equation}
we further normalize $D_+(k,t=t_0)=1$ at $k = 0$.
The linear growth function $D_+$ depends only on time and the magnitude of $\vk$ because  $A(\vk)=A(k)$, which is consistent with the 
rotational invariance of the underlying theory. This does not happen with higher order growth functions since they depend on the manner modes
with different wavelengths interact.

The other solution to Eq.~(\ref{LinearLPTeq}), in general decaying, 
is given by $D_-(k,t)$. It will be used in the Green function associated to the linear operator
$\T - A(k)$: 
\begin{equation}\label{GF}
 (\T - A(k))^{-1} = \int_{t_{in}}^t dt' G(t,t';\vk) = \int_{t_{in}}^t dt' 
 \frac{D_+(k,t)D_-(k,t')-D_+(k,t')D_-(k,t)}{\dot{D}_+(k,t')D_-(k,t')-\typo{\dot{D}_+(k,t')D_-(k,t')}{D_+(k,t')\dot{D}_-(k,t')}}.
\end{equation}
The initial time here is set to $t_{in}$ and a dot means derivative with respect to the time argument. 
In using Eq.~(\ref{GF}) a numerical error may arise if the function at which the operator is applied is not zero at $t_{in}$, 
but can be compensated if
the solution is known at that time, which is the case of models with an early Einstein-de Sitter (EdS) phase, as the one we use below as an example, 
as well as many others found in the literature. 
The alternative is set $t_{in}$ to be very small. For $\Lambda$CDM models there is no $k$-dependence and the growth functions become
$D_- \propto H(t)$ and $D_+ \propto H^{3} (t) \int^{t} a^{-3}(t')H^{-3}(t') dt'$ \cite{1977MNRAS.179..351H}; while for the special case of EdS, 
$D_+ \propto a(t)$ and $D_- \propto a^{-3/2}(t)$. In Fig.~\ref{fig:Dplus} we plot the ratio of MG to $\Lambda$CDM linear growth functions, this is done 
for the F4, F5 and F6 models, and for redshifts $z=0$ and $z=0.5$; the background cosmology is fixed with $\Omega_m=0.281$ and $h = 0.697$, as 
given by WMAP Nine-year results \cite{2013ApJS..208...19H}.

\begin{figure}
	\begin{center}
	\includegraphics[width=3 in]{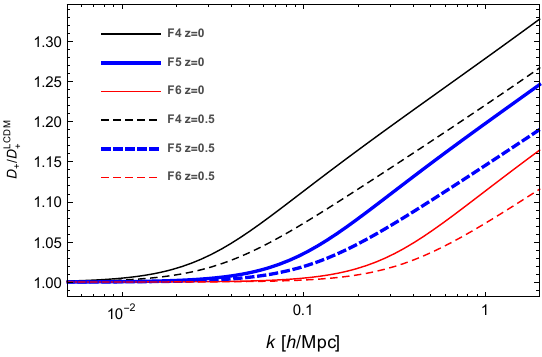}
	\caption{First order growth functions over $\Lambda$CDM growth function. We show the cases of the HS F4, F5 and F6 models and 
	redshifts $z=0$ and $z=0.5$.
	\label{fig:Dplus}}
	\end{center}
\end{figure}

\bigskip

We end this section by writing the scalar field to first order,
\begin{equation}\label{SF1}
\varphi^{(1)}(\vk) = \frac{2A_0}{3 \Pi(k)}  D_+(k,t) \delta^{(1)}_{0}(\vk).
\end{equation}
Since the factor $\frac{2 A_0}{3 \Pi(k)} = 2 a^2\frac{(A(k)-A_0)}{k^2}$ is smaller than unity for typical values of $M_1$, 
it follows that the scalar field perturbations evolve at a slower pace than overdensities.   

\end{section}

\begin{section}{Second order: 2LPT}\label{Sect:2LPT}

In this section we compute second order quantities and develop the 2LPT theory.
The frame-lagging term kernel is obtained from Eq.~(\ref{FLexp}) and the first order fields given in Eqs.~(\ref{1LD}) and (\ref{SF1}),
\begin{align} \label{K2FL}
 \mathcal{K}^{(2)}_\text{FL}(\vk_1, \vk_2) = 2 \frac{(\vk_1 \cdot \vk_2)^2}{k_1^2 k_2^2} (A(k_1) +A(k_2) - 2 A_0) +  
 \frac{\vk_1 \cdot \vk_2}{k_1^2 } (A(k_1)  -  A_0) + \frac{\vk_1 \cdot \vk_2}{k_2^2 } (A(k_2)  -  A_0),
\end{align}
while the self interacting term in Eq.~(\ref{dIexp}) becomes 
\begin{equation}
 \delta I^{(2)} (\vk)  =  \frac{1}{2} \left( \frac{2 A_0}{3} \right)^2 \ikk M_2(\vk_1,\vk_2)
 \frac{D_+(k_1)D_+(k_2)}{\Pi(k_1) \Pi(k_2)} \delta_1 \delta_2.
\end{equation}
Hereafter $\delta_1$ and $\delta_2$ denote the linear density contrasts  with wavenumbers $\vk_1$ and $\vk_2$ 
evaluated at present time. 

A straightforward computation of Eq.~(\ref{eqm}), see Appendix \ref{app::kernels2}, gives the Lagrangian 
displacement to second order:
\begin{equation} \label{LD2order}
\vk_i \Psi^{i (2)}(\vk) = \frac{i }{2} \ikk \frac{3}{7} \left(\bar{D}^{(2)}_a(\vk_1,\vk_2) - 
\bar{D}^{(2)}_b(\vk_1,\vk_2) \frac{(\vk_1 \cdot \vk_2)^2}{k_1^2 k_2^2} - \bar{D}^{(2)}_{\delta I}(\vk_1,\vk_2) + 
\bar{D}^{(2)}_\text{FL}(\vk_1,\vk_2)\right)D_+(k_1)D_+(k_2)\delta_1\delta_2.
\end{equation}
Momentum conservation implies $\vk = \vk_{12} \equiv \vk_1 + \vk_2$, as it is explicit in the Dirac delta function. The growth functions are given by
\begin{align}
  D^{(2)}_a(\vk_1,\vk_2) &=  \big(\T - A(k)\big)^{-1} \big( A(k) D_+(k_1) D_+(k_2) \big) \label{D2a} \\
  D^{(2)}_b(\vk_1,\vk_2) &= \big(\T - A(k)\big)^{-1}  \big( (A(k_1) + A(k_2) - A(k) ) D_+(k_1) D_+(k_2) \big) , \label{D2b} \\
  D^{(2)}_\text{FL}(\vk_1,\vk_2) &=  \big(\T - A(k)\big)^{-1}  \left(  \frac{M_1(k)}{3 \Pi(k)} 
  \mathcal{K}^{(2)}_\text{FL}(\vk_1,\vk_2)   D_+(k_1) D_+(k_2) \right),  \label{D2FL} \\
 D^{(2)}_{\delta I}(\vk_1,\vk_2) &= \big(\T - A(k)\big)^{-1} 
\left(  \left(\frac{2 A_0}{3}\right)^2 \frac{k^2}{a^2}\frac{M_2(\vk_1,\vk_2) D_+(k_1) D_+(k_2)}{6 \Pi(k)\Pi(k_1)\Pi(k_2)} \right),  \label{D2dI} 
\end{align}
and the normalized growth functions are defined as
\begin{align}
 \bar{D}^{(2)}_{a,b,\delta I,\text{FL}}(\vk_1,\vk_2,t) &= \frac{7}{3}\frac{D^{(2)}_{a,b,\delta I,\text{FL}}(\vk_1,\vk_2,t) }{D_+(k_1)D_+(k_2)}. 
\end{align}

The growth functions depend only on three numbers, these can be chosen as $k_1$, $k_2$ and $k_{12}$, 
such that alternatively we can write  
\begin{equation} \label{D2notations}
 \bar{D}^{(2)}(\vk_1,\vk_2)=\bar{D}^{(2)}(k_{12},k_1,k_2).
\end{equation}
Both notations will be used interchangeably.

The linear operator $(\T - A(k))$ is not invertible over its whole domain, and $(\T - A(k))^{-1} h$ is only a 
particular solution to the differential equation $(\T - A(k)) f = h$. The growth functions obtained from 
Eqs.~(\ref{D2a})-(\ref{D2dI}) project out the linear order solution to Eq.~(\ref{LinearLPTeq}), as it is required for being pure second order functions.
On the other hand, numerical computations using differential equations instead, present no disadvantages if 
the initial conditions are carefully chosen.

For dark energy models in which we can neglect dark energy perturbations or MG scale independent models, we have 
$I_2(t) \equiv \bar{D}^{(2)}_a(t) = \bar{D}^{(2)}_b(t)$ (following the notation of \cite{Lee:2014maa}), and 
$\bar{D}^{(2)}_{\delta I} = \bar{D}^{(2)}_\text{FL} = 0$.
For EdS, these reduce to  $I_2(t) = 1$, and for $\Lambda$CDM 
$I_2(t) \simeq  1.01$ at the present time for $\Omega_{m0} \simeq 0.3$.

\bigskip

We can write the divergence of the Lagrangian displacement as 
\begin{equation}\label{LD2order2}
 k_i \Psi^{(2)i}(\vk) =  \frac{i}{2}\ikk D^{(2)} (\vk_1,\vk_2) \delta_1 \delta_2,
\end{equation}
with
\begin{equation} \label{D2whole}
 D^{(2)} (\vk_1,\vk_2) = D^{(2)}_a(\vk_1,\vk_2) - 
D_b^{(2)}(\vk_1,\vk_2) \frac{(\vk_1 \cdot \vk_2)^2}{k_1^2 k_2^2} - D^{(2)}_{\delta I}(\vk_1,\vk_2) + 
D^{(2)}_\text{FL}(\vk_1,\vk_2),
\end{equation}
from which we can read the extension to the LPT second order kernel as
\begin{equation}
 L^{(2)i}(\vk_1,\vk_2) = \frac{k^i}{k^2}  \frac{D^{(2)} (\vk_1,\vk_2)}{D_+(k_1)D_+(k_2)}.
\end{equation}

\bigskip

Consider now the collapse of parallel sheets of matter. In GR it is well known that the equation of motion for the 
Lagrangian displacement becomes linear and the Zel'dovich approximation is the exact solution. This is
because the Newtonian gravitational force, which decays as the inverse of the squared distance, 
becomes a constant \typo{regardless the}{regardless of the} separation of the sheets \cite{McQuinn:2015tva}.
Therefore, the second and higher order Lagrangian displacement kernels are zero. 
This can be deduced from Eq.~(\ref{D2whole}) by setting $D^{(2)}_a=D^{(2)}_b$, 
$D^{(2)}_{\delta I} = D^{(2)}_\text{FL}= 0$, and choosing parallel wavevectors $\vk_1 \parallel \vk_2$, in this situation  
$D^{(2)}_{\text{GR}} (\vk_1,\vk_2) = 0$, as required. In modified theories of gravity the situation is rather different, 
the fifth force between the sheets is \typo{not}{no} longer a constant, and the growth function $D^{(2)} (\vk_1,\vk_2)$ does not vanish since 
$D^{(2)}_a (\vk_1,\vk_2) \neq D^{(2)}_b (\vk_1,\vk_2)$. Nevertheless, at the scales 
the fifth force can be neglected we have to recover the GR result, this should be the case in the large scales limit of $k \rightarrow 0$. Since 
$D_a^{(2)}(k=0, k_1, k_2) \neq D_b^{(2)}(k=0, k_1, k_2)$ and $D^{(2)}_{\delta I}(k=0, k_1, k_2)=0$, the cancellation should be accomplished by the 
frame-lagging. Indeed, the squeezed configuration of triangles formed as $\vk = \vk_1+\vk_2$ 
with $\vk_{12} \simeq 0$, $\vk_1 \simeq \vp$ and $\vk_2 \simeq -\vp$ is a planar collapse; and, 
by using Eqs.~(\ref{K2FL}),(\ref{D2a})-(\ref{D2dI}),(\ref{D2whole}), and the identity $M_1(0) = 3 \Pi(0)$, 
it is easy to see that $D^{(2)} (\vp,-\vp)=0$, recovering GR when $k^2/a^2 \ll M_1$.


As it was mentioned in the Introduction, 2LPT equations in MG have been derived in two (to our knowledge) 
separate works \cite{Valogiannis:2016ane,Winther:2017jof}. In \cite{Valogiannis:2016ane} the results differ with the given here, since 
the authors find  $D_a^{(2)} = D_b^{(2)}$ and, furthermore, they do not seem to consider the frame-lagging. In \cite{Winther:2017jof} their computations 
to second order coincide with ours. In those works the authors use 2LPT for implementations of MG into the COLA code, obtaining very satisfactory
results. These works share the $\Lambda$CDM as a limit at large scales, where 
the 2LPT in COLA implementations is more important. Indeed, to speed-up computations, in Ref.~\cite{Valogiannis:2016ane} 
the 2LPT of $\Lambda$CDM is additionally used into the MG runs, finding also a good agreement with $N$-body simulations.

\begin{figure}
	\begin{center}
		
	\includegraphics[width=3 in]{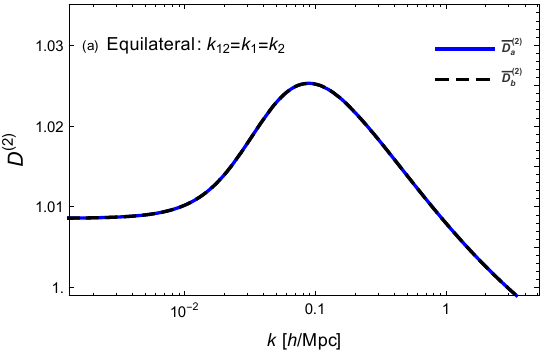} 
		\hspace{0.3cm}
	\includegraphics[width=3 in]{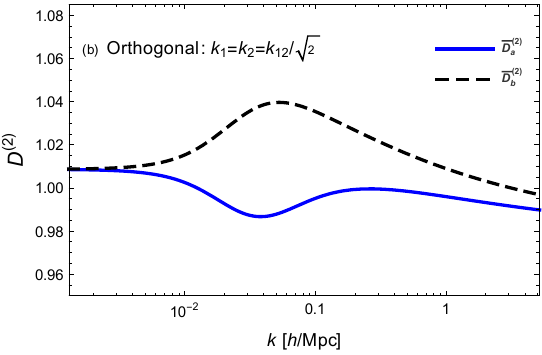}
	\includegraphics[width=3 in]{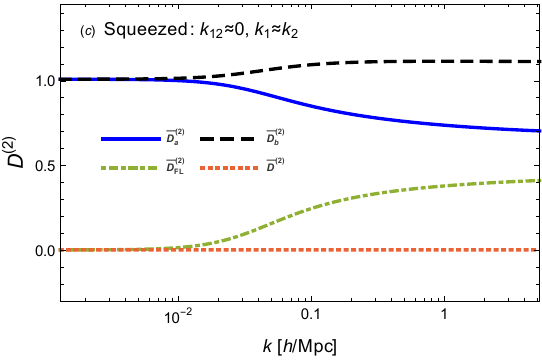} 
		\hspace{0.3cm}
	\includegraphics[width=3 in]{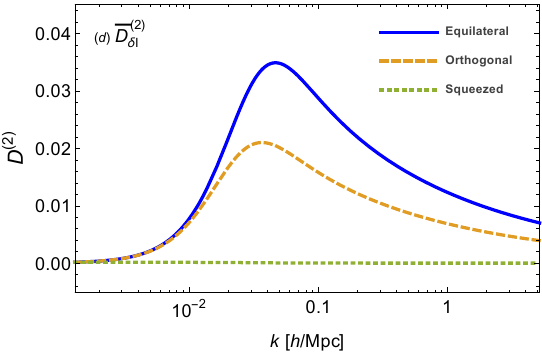} 
	\caption{Second order normalized growth functions $\bar{D}^{(2)}_a$, $\bar{D}^{(2)}_b$,$\bar{D}^{(2)}_{\delta I}$ and $\bar{D}^{(2)}_\text{FL}$,  
	for the F4 model. Different triangular configurations are considered. See text for details. 
	\label{fig:D2}}
	\end{center}
\end{figure}

In Fig.\ref{fig:D2} we plot second order growth functions for the F4 case\footnote{The effects of F5, 
and F6 are qualitatively very similar but are shifted toward higher values of $k$.}
and we choose three different triangular configurations: (a) equilateral 
for which $k_{12} = k_1 = k_2 $; (b) orthogonal with $k_{12} =  \sqrt{2}k_1  =\sqrt{2}k_2$; and (c) squeezed, $k_{12}\simeq 0$, $k_1 \simeq k_2$. We can read from 
the equations for the growth functions that in the case of equilateral configuration we have $\bar{D}_a^{(2)}=\bar{D}_b^{(2)}$. 
The squeezed case corresponds to
the limit of large scales where the theory reduces to $\Lambda$CDM, we can see from panel (c) in Fig.~\ref{fig:D2} that actually by considering
the combination $ \bar{D}^{(2)}_a - \bar{D}^{(2)}_b + \bar{D}^{(2)}_\text{FL}$ (dotted red line in panel (c)) 
in this configuration we get $D^{(2)}=0$, as was shown analytically above.
The frame-lagging second order growth function for equilateral configuration is exactly zero and for the orthogonal is very similar in shape to the 
squeezed one,  but reaching smaller values. In panel (d) we show the screening normalized growth functions for the three 
considered triangular configurations. We make note
that the functions $\bar{D}_a^{(2)}$ and $\bar{D}_b^{(2)}$ do not tend to 1 as $k$ goes to zero, the EdS value, 
instead their limits are $\sim 1.009$, the $I_2(t_0)$ value for the chosen cosmological parameters in the $\Lambda$CDM model. 

\bigskip

Now, we calculate the matter and scalar field perturbation to second order. From Eq.~(\ref{deltax}), 
$\tilde{\delta}^{(2)}(\vk) =  [-\Psi_{i,i}^{(2)} + \frac{1}{2} (\Psi_{i,i}^{(1)}\Psi_{j,j}^{(1)} + \Psi_{i,j}^{(1)}\Psi_{j,i}^{(1)})](\vk)$, then
\begin{equation}
\tilde{\delta}^{(2)}(\vk) = \frac{1}{2} \ikk \left( D^{(2)} (\vk_1,\vk_2) 
+ \left(1 + \frac{(\vk_1\cdot\vk_2)^2}{k_1^2 k_2^2}\right) D_+(k_1) D_+(k_2) \right) \delta_1 \delta_2,
\end{equation}
and the scalar field becomes
\begin{equation}\label{BD2order}
 \varphi^{(2)}(\vk) = \frac{2A_0}{3 \Pi(k)} \frac{1}{2} \ikk D^{(2)}_{\varphi}  (\vk_1,\vk_2) \delta_1 \delta_2,
\end{equation}
with
\begin{align}
 D^{(2)}_{\varphi}  (\vk_1,\vk_2) &= D^{(2)} (\vk_1,\vk_2) 
+ \left(1 + \frac{(\vk_1\cdot\vk_2)^2}{k_1^2 k_2^2}\right) D_+(k_1) D_+(k_2) \nonumber\\*
& - \frac{2 A_0}{3}  \frac{M_2(\vk_1,\vk_2)+ J_\text{FL}(\vk_1,\vk_2) (3 + 2\obd)}{3 \Pi(k_1)\Pi(k_2)}D_+(k_1) D_+(k_2).
\end{align}
We shall use the above perturbations in the following section to find the third order Lagrangian displacement field.

\end{section}

\begin{section}{Third order}\label{Sect:3LPT}
 
In this section we compute the expansion of the relevant fields up to third order. Using Eqs. (\ref{FLexp}), (\ref{LD2order2}) 
and (\ref{BD2order}), the frame-lagging kernel is
\begin{align}\label{FLK3}
 \mathcal{K}^{(3)}_\text{FL}(\vk_1,\vk_2,\vk_3) &= 
 3  \left( 2 \frac{(\vk_1 \cdot \vk_{23})^2}{k_1^2 k_{23}^2} +  \frac{\vk_1 \cdot \vk_{23}}{k_1^2 } \right) 
 \frac{D^{(2)}(\vk_2,\vk_3) }{D_+(k_2)D_+(k_3)}  (A(k_1)-A_0)\nonumber\\*
 &+ 3  \left( 2 \frac{(\vk_1 \cdot \vk_{23})^2}{k_1^2 k_{23}^2} +  \frac{\vk_1 \cdot \vk_{23}}{k_{23}^2 } \right) 
 \frac{D^{(2)}_\varphi (\vk_2,\vk_3)}{D_+(k_2)D_+(k_3)}  (A(k_{23})-A_0) \nonumber\\*
 &+ 6 \left( 3 \frac{(\vk_1\cdot\vk_2)(\vk_2\cdot\vk_3)(\vk_3\cdot\vk_1)}{k_2^2k_3^2k_1^2} 
 + 2 \frac{(\vk_2\cdot\vk_3)^2(\vk_3\cdot\vk_1)}{k_2^2k_3^2k_1^2} 
 +  \frac{(\vk_2\cdot\vk_3)(\vk_3\cdot\vk_1)}{k_3^2k_1^2} \right)  (A(k_1)-A_0).
\end{align}
The third order screening gives
\begin{equation}
 \delta I^{(3)} = \frac{1}{6} \ikkk \mathcal{K}^{(3)}_{\delta I}(\vk_1,\vk_2,\vk_3) D_+(k_1)D_+(k_2)D_+(k_3) \delta_1 \delta_2  \delta_3,
\end{equation}
with kernel
\begin{align}
 \mathcal{K}^{(3)}_{\delta I}(\vk_1,\vk_2,\vk_3) &= 
 \left( \frac{2A_0}{3}\right)^2 \frac{3 M_2(\vk_1,\vk_{23})}{\Pi(k_1) \Pi(k_{23})}  \left(1 + \frac{(\vk_2\cdot\typo{\vk_2}{\vk_3})^2}{k_2^2k_3^2} 
 + \frac{D^{(2)}(\vk_2,\vk_3)}{D_+(k_2)D_+(k_3)} \right) \nonumber\\*
&+ \left( \frac{2A_0}{3}\right)^3 \left( \frac{M_3(k_1,k_2,k_3)}{\Pi(k_1)\Pi(k_2)\Pi(k_3)} -
\frac{ M_2(\vk_1,\vk_{23})(M_2(\vk_2,\vk_3)+J^{(2)}_{FL}(\vk_2,\vk_3)(3+ 2 w_\text{BD}))}{\Pi(k_{23})\Pi(k_1)\Pi(k_2)\Pi(k_3)} \right).
\end{align}
A straightforward but lengthy computation, see Appendix \ref{app::kernels3}, leads to
\begin{align} \label{LD3order}
  k_i \Psi^{(3)i}(\vk) &= \frac{i}{6} \ikkk  D_{+}(k_1)D_{+}(k_2)D_{+}(k_3) \delta_1 \delta_2 \delta_3
    \Bigg\{ \frac{5}{7} \left( \bar{D}^{(3)}_{A} -\bar{D}^{(3)}_{B}
              \frac{(\vk_2 \cdot \vk_3)^2}{k^2_2 \typo{k^3_2}{k^2_3}} + \bar{D}^{(3)}_{CTa} \right) 
             \left( 1-  \frac{(\vk_1 \cdot \vk_{23})^2}{k_1^2 k_{23}^2} \right)  \nonumber\\*
    &  \quad\qquad -\frac{1}{3} \left( \,  \bar{D}^{(3)}_{C}  -3 \bar{D}^{(3)}_{D} \frac{(\vk_2 \cdot \vk_3)^2}{k^2_2 \typo{k^3_2}{k^2_3}} 
     + 2 \bar{D}^{(3)}_{E} \frac{(\vk_1 \cdot \vk_2)(\vk_2 \cdot \vk_3)(\vk_3 \cdot \vk_1)}{k_1^2 k^2_2 k^2_3}  + \bar{D}^{(3)}_{CTb}\, \right)
     - \bar{D}^{(3)}_{\delta I} + \bar{D}^{(3)}_\text{FL} \Bigg\},                             
 \end{align} 
with normalized growth functions 
\begin{align}
 \bar{D}^{(3)}_{A,B,CTa}(\vk_1,\vk_2,\vk_3) &= \frac{7}{5}  \frac{D^{(3)}_{A,B,CTa}(\vk_1,\vk_2,\vk_3)}{D_{+}(k_1)D_{+}(k_2)D_{+}(k_3)}, \\
 \bar{D}^{(3)}_{C,D,E,\delta I, \text{FL},CTb}(\vk_1,\vk_2,\vk_3) 
 &= \frac{  D^{(3)}_{C,D,E,\delta I, \text{FL},CTb}(\vk_1,\vk_2,\vk_3)}{D_{+}(k_1)D_{+}(k_2)D_{+}(k_3)},
\end{align} 
and growth functions
\begin{align} 
 D^{(3)}_A &= \left(\T - A(k)\right)^{-1} \left( 3 D_+(k_1) \big(A(k_1) + \T - A(k)\big)D^{(2)}_a(\vk_2,\vk_3) \right), \\
 D^{(3)}_B &= \left(\T - A(k)\right)^{-1} \left( 3 D_+(k_1) \big(A(k_1) + \T - A(k)\big)D^{(2)}_b(\vk_2,\vk_3) \right), \\
 D^{(3)}_{CTa} &= \left(\T - A(k)\right)^{-1} \left(3 D_+(k_1)\big(A(k_1) + \T - A(k)\big) 
                  \big( D^{(2)}_\text{FL}(\vk_2,\vk_3)-D^{(2)}_{\delta I}(\vk_2,\vk_3) \big) \right),\\ 
 D^{(3)}_C &= \left(\T - A(k)\right)^{-1} \left( 9 D_+(k_1) \big(A(k_1) + \T - 2 A(k)\big)D^{(2)}_a(\vk_2,\vk_3)
                                            -3 A(k) D_+(k_1)D_+(k_2)D_+(k_3) \right), \\
 D^{(3)}_D &= \left(\T - A(k)\right)^{-1} \left( 3 D_+(k_1) \big(A(k_1) + \T - 2 A(k)\big)D^{(2)}_b(\vk_2,\vk_3)
                                            +3 A(k) D_+(k_1)D_+(k_2)D_+(k_3) \right), \\
 D^{(3)}_E &= \left(\T - A(k)\right)^{-1} \Big( 3 \big(3 A(k_1) - A(k)\big)D_+(k_1)D_+(k_2)D_+(k_3) \Big), \\ 
 D^{(3)}_{CTb} &= \left(\T - A(k)\right)^{-1} \left(9 D_+(k_1)\big(A(k_1) + \T -2 A(k)\big)
                   \big( D^{(2)}_\text{FL}(\vk_2,\vk_3) -D^{(2)}_{\delta I}(\vk_2,\vk_3) \big) \right), \\
 D^{(3)}_{\delta I} &= \left(\T - A(k)\right)^{-1} \left( \frac{k^2/a^2}{6 \Pi(k)}  
 \mathcal{K}^{(3)}_{\delta I}(\vk_1,\vk_2,\vk_3) D_+(k_1)D_+(k_2)D_+(k_3) \right),\\ 
 D^{(3)}_\text{FL} &= \left(\T - A(k)\right)^{-1} \left( \frac{M_1}{3 \Pi(k)}  \mathcal{K}^{(3)}_\text{FL}(\vk_1,\vk_2,\vk_3) D_+(k_1)D_+(k_2)D_+(k_3) \right).
\end{align}  
The growth functions depend on three wavevectors, but these are constrained to form a quadrilateral $\vk = \vk_1 + \vk_2 + \vk_3$; 
thus, they depend only on 6 numbers for a given $\vk$. Furthermore, the relevant configurations  for these quadrilaterals 
in 2-point statistics are
the so-called double squeezed in which $\vk = - \vk_1$ and $\vp \equiv \vk_3 = -\vk_2$, which left us with only 3 degrees of freedom,  
highly simplifying the analysis.

For the $\Lambda$CDM model we have $D^{(3)}_{A} = D^{(3)}_{B} =(\T - A_0)^{-1}(\frac{9}{7} D_+ \T D_+^2)$, and
since $\T D_+^2 =2 D_+ \T D_+ + 2 \dot{D}_+^2$, we get 
$D^{(3)}_{A,B}= \frac{5}{7} 6 (\T - A_0)^{-1}(\frac{3}{2} \Omega_m H^2 D_+^3 (\frac{3}{5} + \frac{2}{5} f^2/\Omega_m))$, with 
$f = d \ln \delta_L / d \ln a$ the growth factor of linear matter perturbations. For
$f = \Omega_m^{1/2}$ we obtain the result for EdS, $D^{(3)}_{A,B} = \frac{5}{7}  D_+^3$, meaning that the normalized growth functions 
$\bar{D}^{(3)}_{A,B}$ are exactly 1. 
While for $\Lambda$CDM we get $\bar{D}^{(3)}_{A,B}(t_0) \simeq 1.02$ for $\Omega_{m0} \simeq 0.3$. Analogously, we obtain
$D^{(3)}_{C,D,E} = 6 (\T - A_0)^{-1}(\frac{3}{2} \Omega_m H^2 D_+^3 )$, which reduces to $D^{(3)}_{C,D,E} = D_+^3$ for EdS, while for 
$\Lambda$CDM we find $\bar{D}^{(3)}_{C,D,E}  \simeq 1.02$ at the present time. 
This analysis shows that the third order displacement field given by Eq.~(\ref{LD3order}) reduces
to the standard case when the theory is scale independent.

We can write the third order Lagrangian displacement kernel as
\begin{equation}\label{3LDkernel}
 L^{(3)i}(\vk_1,\vk_2,\vk_3) = \typo{i\frac{k^i}{k^2}}{\frac{k^i}{k^2}} \frac{D^{(3)}(\vk_1,\vk_2,\vk_3)}{D_{+}(k_1)D_{+}(k_2)D_{+}(k_3)},
\end{equation}
with
\begin{align} 
 D^{(3)}(\vk_1,\vk_2,\vk_3) &= \left( D^{(3)}_{A} -D^{(3)}_{B}\frac{(\vk_2 \cdot \vk_3)^2}{k^2_2 \typo{k^3_2}{k^2_3}} + D^{(3)}_{CTa} \right) 
             \left( 1-  \frac{(\vk_1 \cdot \vk_{23})^2}{k_1^2 k_{23}^2} \right)  \nonumber\\
    &  -\frac{1}{3} \left( \,  D^{(3)}_{C}  -3 D^{(3)}_{D} \frac{(\vk_2 \cdot \vk_3)^2}{k^2_2 \typo{k^3_2}{k^2_3}} 
     + 2 D^{(3)}_{E} \frac{(\vk_1 \cdot \vk_2)(\vk_2 \cdot \vk_3)(\vk_3 \cdot \vk_1)}{k_1^2 k^2_2 k^2_3}  + D^{(3)}_{CTb}\, \right)
     - D^{(3)}_{\delta I} + D^{(3)}_\text{FL}.
\end{align}
In Eq.(\ref{3LDkernel}) we omit to write a transverse part because it cancels when it is contracted with $\vk$.

\bigskip

In 1-loop statistics, the third order growth function should be symmetrized 
by summing over all permutations ---contrary to cyclic permutations which are 
sufficient in $\Lambda$CDM. By doing this, we obtain
\begin{align} \label{D3symm}
 &  D^{(3)symm} (\vk,-\vp,\vp)=  
 \left(\T - A(k)\right)^{-1} \Bigg\{  D_+(p)\left(A(p) + \T - A(k)\right) D^{(2)}(\vp,\vk) 
                \left( 1- \frac{(\vp \cdot (\vk + \vp))^2}{p^2 |\vp+\vk|^2}\right) \nonumber\\* 
&   - D_+(p)\left(A(p) + A(|\vk + \vp|) - 2A(k)\right) D^{(2)}(\vp,\vk)  
               +  \left(2A(k) -  A(p) - A(|\vk + \vp|) \right) D_+(k) D_+^2(p) \frac{(\vk \cdot \vp)^2}{k^2p^2}\nonumber\\*
&   - \left( A(|\vk + \vp|) -  A(k) \right) D_+(k) D_+^2(p) 
               -  \left( \frac{M_1(\vk + \vp)}{3\Pi(|\vk + \vp|)}\mathcal{K}^{(2)}_\text{FL}(\vp,\vk) 
               -  \left(\frac{2 A_0}{3}\right)^2  \frac{M_2(\vp,\vk) |\vk + \vp|^2/a^2}{6\Pi(|\vk + \vp|)\Pi(k)\Pi(p)} \right)  D_+(k) D_+^2(p) \nonumber\\*
& + \frac{M_1(k)}{3 \Pi(k)} \Bigg[ \left( \typo{}{2}\frac{(\vp \cdot (\vk + \vp))^2}{p^2 |\vp+\vk|^2}  - \frac{\vp\cdot(\vk+\vp)}{p^2}\right)
    (A(p) - A_0) D^{(2)}(\vp,\vk) D_+(p) +\left( \typo{}{2}\frac{(\vp \cdot (\vk + \vp))^2}{p^2 |\vp+\vk|^2}  - \frac{\vp\cdot(\vk+\vp)}{|\vk+\vp|^2}\right)  \nonumber\\*
&   \times (A(|\vk+\vp|) - A_0) D^{(2)}_\varphi(\vp,\vk) D_+(p) + 3 \frac{(\vk \cdot \vp)^2}{k^2 p^2} 
    \left( A(k) + A(p) - 2 A_0 \right) D_+(k) D_+^2(p)   \Bigg] \nonumber\\*
&    - \frac{1}{2} \frac{k^2/a^2}{6\Pi(k)} \mathcal{K}_{\delta I}^{(3)symm} (\vk,-\vp,\vp) D_+(k) D_+^2(p)
 \,\,\Bigg\} \quad + \quad (\, \vp \rightarrow -\vp \,), 
\end{align}
where in writing ``$ (\, \vp \rightarrow -\vp \,) $'', we assume that $M_2(\vp,\vk)$ is symmetric in $\vk$ and $\vp$,
otherwise it should be symmetrized.
It is far from obvious that Eq.~(\ref{LD3order}) goes to its $\Lambda$CDM form as $\vk$ goes to zero. But by taking the limit case $k=0$ 
to Eq.~(\ref{D3symm}), several cancellations lead to $D^{(3)symm} (0,-\vp,\vp) = 0$, as it should be the case
since double squeezed configurations reduce to the 1-dimensional collapse in the limit of $k\rightarrow 0$. 
The expression for $\mathcal{K}_{\delta I}^{(3)symm}$ is somewhat large and not necessary 
for recovering GR, thus we do not write it here.

\bigskip

We finish this section by making some comments on the success of the Zel'dovich approximation at the large scales; for further discussion 
see \cite{Tassev:2013pn,Whi14,McQuinn:2015tva,Cusin:2016zvu}. 
The second order Lagrangian displacement trivially reduces to 
a planar (one dimensional) collapse when $k\rightarrow 0$ simply because only two modes with $\vk_1$ and $\vk_2$ 
interact and by conservation of momentum
 $\vk_1 \rightarrow - \vk_2$. For the third order Lagrangian displacement 
 the wavelength vectors should form a quadrilateral $\vk = \vk_1 + \vk_2 + \vk_3$, again due to 
 momentum conservation; and moreover, statistics as the power spectrum or the correlation function rely 
 on double-squeezed configurations which also reduce to planar collapse
in the case $k\rightarrow 0$. Thus, 2-point statistics at large scales essentially probe planar collapses, for which 
Zel'dovich approximation is exact.

\end{section}

\begin{section}{Lagrangian displacements 2- and 3-point functions}\label{Sect:LDstats}

Lagrangian displacement power spectra and bispectra are 2- and 3-rank tensors, of the form
$\langle \Psi_i(k_1) \Psi_j(k_2) \rangle$ and $\langle \Psi_i(k_1) \Psi_j(k_2) \Psi_k(k_3)\rangle$, that we
contract with related momenta to construct scalars. In this section we are interested in 
those combinations that are necessary for matter statistics at 1-loop. These special
combinations are defined in Appendix \ref{app::correlations}, Eqs.(\ref{defQ1})-(\ref{defR2}). 
Straightforward calculations using the Lagrangian displacements up to third order yield  
\begin{align}
 Q_1(k) &=  \frac{k^3}{4 \pi^2} \int_0^\infty dr r^2 P_L(kr) \int^{1}_{-1} dx P_L(k\sqrt{1+r^2 - 2 r x})
\left( \bar{D}^{(2)}_a - \bar{D}^{(2)}_b \left( \frac{x^2 + r^2 - 2 r x}{1 + r^2 - 2 r x}\right)  
- \bar{D}^{(2)}_{\delta I} +  \bar{D}^{(2)}_\text{FL}  \right)^2, \label{Q1f} \\ 
 Q_2(k) &=\frac{k^3}{4 \pi^2} \int_0^\infty dr P_L(kr)
\int_{-1}^1 dx P_L(k\sqrt{1+r^2 -2 r x})  \frac{rx(1-rx)}{1+r^2 -2 r x}
\left( \bar{D}^{(2)}_a - \bar{D}^{(2)}_b \frac{(x-r)^2}{1+r^2 -2 r x} 
+  \bar{D}^{(2)}_\text{FL} -  \bar{D}^{(2)}_{\delta I} \right),  \label{Q2f} \\ 
 Q_3(k) &= \frac{k^3}{4 \pi^2}   \int_0^\infty dr  P_L(kr)  \int_{-1}^1 dx \frac{x^2(1-rx)^2}{(1+r^2 -2 r x)^2} P_L(k\sqrt{1+r^2 -2 r x}), \label{Q3f} \\ 
 R_1(k) &= \typo{}{\frac{21}{10}}\frac{k^3}{4 \pi^2} P_L(k) \int_0^\infty dr \,\typo{}{r^2} P_L(k r)  \int^{1}_{-1} dx \frac{D^{(3)symm}(\vk,-\vp,\vp)}{D_{+}(k)D_{+}^2(p)}, \label{R1f} \\ 
  R_2(k) &=  \frac{k^3}{4 \pi^2} P_L(k)    \int_0^\infty dr P_L(kr) \int_{-1}^1 dx  
\frac{rx(1-rx)}{1+r^2 -2 r x}  \left( \bar{D}^{(2)}_a - \bar{D}^{(2)}_b x^2 
+  \bar{D}^{(2)}_\text{FL} -  \bar{D}^{(2)}_{\delta I} \right), \label{R2f}
\end{align}
with $r=p/k$ and $x= \hat{\vk}\cdot\hat{\vp}$.
The normalized growth functions in Eqs.~(\ref{Q1f}) and (\ref{Q2f}) are evaluated as (using the notation of Eq~(\ref{D2notations}))
\begin{align}
D^{(2)}(\vp,\vk-\vp) &= D^{(2)}(k,k r, k \sqrt{1+r^2 - 2 x r}) \qquad \text{for} \quad Q_1 \,\, \text{and} \,\, Q_2, \label{Q1Q2Ev}
\end{align}
and in Eq.~(\ref{R2f}):
\begin{align}
D^{(2)}(\vk,-\vp)    &= D^{(2)}( k \sqrt{1+r^2 - 2 x r},k,k r) \qquad \text{for} \quad R_2. \label{R2Ev}
\end{align}
We emphasize that in the notation of the left hand side of the above equation, the growth functions are symmetric in their arguments. 
The evaluation of the third order growth function was already given above, in Eq.~(\ref{D3symm}). 
In this section we consider the F4 model, since their MG effects are more evident than in F5 and F6, but the qualitative features are not 
affected.

\begin{figure}
	\begin{center}
	\includegraphics[width=3 in]{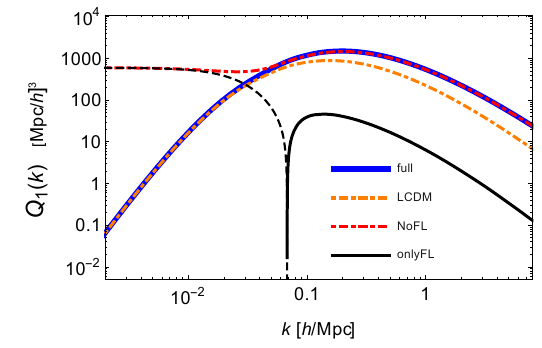}
	\includegraphics[width=3 in]{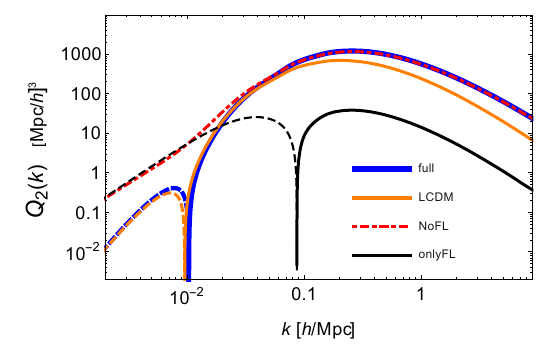}
	\caption{Functions $Q_1(k)$ and $Q_2(k)$ for F4 model: the dot-dashed red curves do not contain
	frame-lagging terms, the black curves have pure frame-lagging contributions (with the dashing showing the
	negative), and the solid blue is the full curve (with dashing for the negative in $Q_2$), the dotted orange is the $\Lambda$CDM. 
	\label{fig:Qs}}
	\end{center}
\end{figure}

In Fig.~\ref{fig:Qs} we show plots for the $Q_1$ and $Q_2$ functions considering their different contributions. 
Terms containing only frame-lagging are shown with black curves,
while the red dot-dashed have no frame-lagging contributions, 
the sum of both gives the full $Q_1$ or $Q_2$ functions (blue curves), and the orange curves are for the
$\Lambda$CDM model. 
For $Q_1$, the cancellation due to the frame-lagging is about one part in a ten thousand at $k=0.001$, 
showing the importance of these terms at large scales. For $Q_2$ the cancellation is about one order of magnitude
at the same scale.
Analogously, in Fig.~\ref{fig:Rs} we plot the $R_1$ and $R_2$ functions. In these cases the cancellations are not 
sufficient to drive these functions to their $\Lambda$CDM counterparts. More detailed inspections of their large 
scales behavior reveals the reason, as we show below.

To emphasize the importance of frame-lagging terms we consider first the $Q_1$ function. 
By letting the external momentum $k$ goes to zero in Eq.~(\ref{Q1f}) we have
\begin{align}
Q_1(k \rightarrow 0) &=  \frac{1}{4\pi^2}\int_0^\infty dp\, p^2 P_L^2(p) 
 \int_{-1}^{1} d x \left(\mathcal{A}-\mathcal{B}\frac{k^2 x^2 + p^2 - 2 k p x}{k^2 +p^2-2kpx} \right)^2 , 
\end{align}
with $\mathcal{A} = \bar{D}^{(2)}_a(\vp,-\vp) + \bar{D}^{(2)}_\text{FL}(\vp,-\vp)$ and
$\mathcal{B} = \bar{D}^{(2)}_b(\vp,-\vp)$. The $\bar{D}^{(2)}_{\delta I}$ growth function vanishes for this configuration, otherwise 
it would contribute to the function $\mathcal{A}$. 
Expanding the integrand about $k=0$ and discarding terms with odd powers in $x$ since they become zero after the angular integration, we get
\begin{align} \label{expintQ1}
\left(\mathcal{A}-\mathcal{B}\frac{k^2 x^2 + p^2 - 2 k p x}{k^2 +p^2-2kpx} \right)^2  &=
(\mathcal{A}-\mathcal{B})^2 + 2 \mathcal{B}(\mathcal{A}-\mathcal{B})(1-x^2) \frac{k^2}{p^2} \nonumber\\
& + \big(\mathcal{B}^2(1-x^2)^2 + 2 \mathcal{B}(\mathcal{A}-\mathcal{B})(-1\typo{-}{+}5x^2- 4 x^4) \big) \frac{k^4}{p^4} + \mathcal{O}\left( \frac{k^6}{p^6}\right).
\end{align}
These squeezed configurations survey large scales and then should reduce to GR. We 
explicitly showed in Sect.~\ref{Sect:2LPT} that 
$\mathcal{A}-\mathcal{B} = \bar{D}^{(2)}_a(\vp,-\vp) - \bar{D}^{(2)}_b(\vp,-\vp)  + \bar{D}^{(2)}_\text{FL}(\vp,-\vp) =0$, thanks to
the frame-lagging. By this reason
the leading term that survives in Eq.~(\ref{expintQ1}) is of order $\mathcal{O}((k/p)^4)$, yielding
\begin{align}\label{Q1kzero}
Q_1(k \rightarrow 0)&= \frac{4 k^4}{15 \pi^2}\int_0^\infty dp \frac{P_L^2(p)}{p^2}  (\bar{D}^{(2)}_b(\vp,-\vp))^2 
   = C_{Q_1}  \frac{4 k^4}{15 \pi^2}\int_0^\infty dp \frac{P_L^2(p)}{p^2}. 
\end{align}
In panel (D) of Fig.~\ref{fig:D2} we plotted $\bar{D}^{(2)}_b(\vp,-\vp)$. It starts to depart from GR value early, at 
yet linear scales, but the integrand above is quickly damped by the $p^{-2}$ factor, thus the net effect is small. 
For F4 we find $C_{Q_1}\simeq 1.03$, for $\Lambda$CDM $C_{Q_1} = I^{2}_2(t=t_0) \simeq 1.02$, and for
EdS $C_{Q_1}$ is exactly one. We note that there
is a residual $k$ dependence in $C_{\typo{Q_2}{Q_1}}$ since the external momentum takes small, but non-zero values. On the other hand, if we neglect 
the frame-lagging, then $\mathcal{A}-\mathcal{B} \neq 0$ and the leading term in the expansion of Eq.~(\ref{expintQ1}) is 
of order zero and we get
\begin{equation}
Q_1^\text{NoFL}(k \rightarrow 0) =  \frac{1}{4 \pi^2}\int_0^\infty dp \, p^2 P_L^2(p)  \big(\bar{D}^{(2)}_a(-\vp,\vp)-\bar{D}^{(2)}_b(-\vp,\vp)\big)^2  
\end{equation}
This explains that in Fig.~\ref{fig:Qs} the function $Q_1$ without frame-lagging approaches a constant for low-$k$, and due to the $p^2$ term the function
takes large values.

\begin{figure}
	\begin{center}
	\includegraphics[width=3 in]{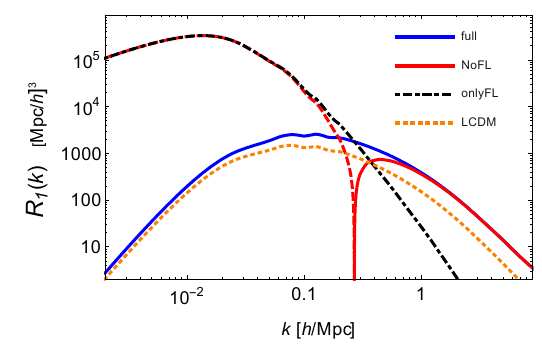}
	\includegraphics[width=3 in]{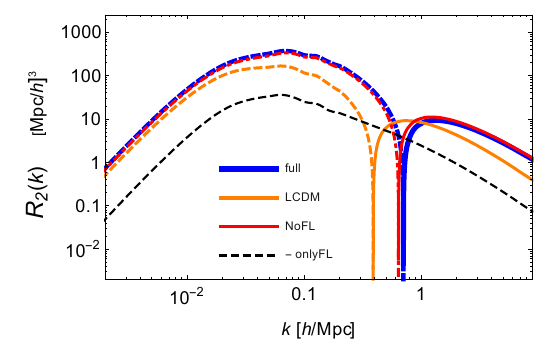}
	\caption{\emph{ Left panel:} $R_1(k)$ function: the solid red curve does not contain
	frame-lagging terms (with the dashing showing its negative); the dot-dashed black curve has pure frame-lagging contributions; 
	the solid blue is the full curve; and the dotted orange is the $\Lambda$CDM. \emph{Right panel:} $R_2(k)$ function: 
	the solid red curve does not contain frame-lagging terms (with the dashing showing its negative); dashed black is the negative
	of pure frame-lagging contributions; solid blue is the full $R_2$ curve  (with the dashing showing its negative); and solid orange
	is for $\Lambda$CDM. 
	\label{fig:Rs}}
	\end{center}
\end{figure}

Analogous calculations for $Q_2$ leads to 
\begin{align}\label{Q2kzero}
Q_2(k \rightarrow 0)&= - \frac{ k^4}{15 \pi^2}\int_0^\infty dp \frac{P_L^2(p)}{p^2}  \bar{D}^{(2)}_b(-\vp,\vp) 
   =- C_{Q_2}  \frac{ k^4}{15 \pi^2}\int_0^\infty dp \frac{P_L^2(p)}{p^2}, 
\end{align}
while without the frame-lagging one would obtain $Q_2^{\typo{\text{FL}}{\text{NoFL}}}(k \rightarrow 0) \propto k^2 \int dp P^2_L(p)$.
The other loop functions yield
\begin{align}
R_1(k \rightarrow 0) &=  C_{R_1}  \frac{4 \typo{k^4}{k^2}}{15 \pi^2} P_L(k) \int_0^\infty dp P_L(p), \label{R1kzero}\\
R_2(k \rightarrow 0) &=  - C_{R_2}  \frac{ \typo{k^4}{k^2}}{15 \pi^2} P_L(k) \int_0^\infty dp P_L(p). \label{R2kzero}
\end{align}
The difference in the derivation of $R_2$ is that the second order growth functions are evaluated
as $\bar{D}^{(2)}(0,-\vp)$, for which frame-lagging terms give small negative contributions, as observed in Fig.~\ref{fig:Rs}.
The case of $R_1$ was guessed and then tested numerically. 
We have checked  the
goodness of these approximations numerically, showing that all the $C_X$ functions are close to unity, but more relevant for this discussion is that 
they depend very weakly on $k$. Therefore, the $Q$ and $R$ functions have the same $k$-dependence at
large scales in MG and GR.

From Eqs.~(\ref{Q1kzero}) and (\ref{Q2kzero}), we now notice that the large scales modes of $Q_1$ and $Q_2$ functions receive negligible contributions from small scales, 
making them indistinguishable for MG models that reduce to $\Lambda$CDM at those scales, as depicted in Fig.~\ref{fig:Qs}. On the other hand,
the $R_1$ and $R_2$ functions at {\it low}-$k$ (see Eqs.~(\ref{R1kzero}) and (\ref{R2kzero})) reduce to the linear power spectrum times the variance
of the Lagrangian displacements, and as a result they take larger values than to those of $\Lambda$CDM, as shown in Fig.~\ref{fig:Rs}.

The analysis of the $Q_3$ function is simpler because it derives from products of linear displacement fields, thus it has the same form
in MG and GR. This function is not necessary for LPT matter 
statistics, but it is required for computing the SPT power spectrum, as explained in Appendix \ref{app::correlations}. Its large scale limit is
\begin{equation}
 Q_3(k \rightarrow 0) =  \frac{ k^4}{10 \pi^2} \int_0^\infty dp \frac{P_L^2(p)}{p^2}.
\end{equation}

\bigskip

Now, for scale invariant universes with $P_L(p) \propto p^n$ as we consider in the following, we fix the external $k$ momentum
and let the internal momentum $p$ goes to infinity. The equations derived above in the limit $k\rightarrow 0$ can do this job for power law power spectra,
and in this way we observe the $Q$ functions
have UV divergences for $n\geq 1/2$ and the $R$ functions for $n \geq -1$;
these are the same UV divergences that have the $P_{22}$ and $P_{13}$ pieces of the SPT power spectrum, respectively. We notice that
if frame-lagging is not considered, UV divergences for $Q_1$ are present for $n\geq -3/2$, while  for $\typo{Q_1}{Q_2}$ when
$n>-1/2$.

\bigskip

On the other hand, IR divergences may show up both at $\vp \rightarrow 0$ and $\vp \rightarrow \vk$. 
In Ref.~\cite{Carrasco:2013sva} $P_{22}$ is written in such a way both divergences appear at 
momentum zero. We follow that work and define
\begin{equation}
  q_3(\vk,\vp)  = \frac{(\vk \cdot \vp)^2}{p^4}\frac{(\vk \cdot (\vk-\vp))^2}{|\vk-\vp|^4}P_L(p) P_L (|\vk-\vp|).
\end{equation}
From Eqs.~(\ref{c11}) and (\ref{defQ3}), $Q_3$ can be written as
\begin{align}
 Q_3(k)  &=  \int \Dk{p} q_3(\vk,\vp)  =  \underset{p<|\vk -\vp|}{\int} \Dk{p} q_3(\vk,\vp)  
           +\underset{p>|\vk -\vp|}{\int} \Dk{p} q_3(\vk,\vp)  \nonumber\\
        &= \underset{p<|\vk -\vp|}{\int} \Dk{p} q_3(\vk,\vp)
           +\underset{\tilde{p}<|\vk - \mathbf{\tilde{p}}|}{\int} \Dk{\tilde{p}} q_3(\vk,\vk - \mathbf{\tilde{p}})  \nonumber\\
        &=  2 \!\!\!\underset{p<|\vk -\vp|}{\int} \Dk{p} q_3(\vk,\vp).
\end{align}
In the second equality we have split the region of integration in two pieces separated by the $p\sim k$ divergence. In the 
second integral of the third equality we redefined the variable $\vp = \vk - \mathbf{\tilde{p}}$. In the last we use the 
symmetry $q_3(\vk,\vp) = q_3(\vk,\vk-\vp)$.  In this way, $Q_3(k)$ has IR divergences only for $\vp \rightarrow 0$.
Fixing the external momentum, a standard computation leads to write, with a little abuse of notation,
\begin{equation}\label{Q3IR}
 Q_3^\text{IR}= \frac{k^2}{3\pi^2} P_L(k) \int_0^{k} dp P_L(p) \left(1 + \frac{p^2}{k^2} + \mathcal{O}\left( \frac{p^4}{k^4}\right) \right), 
\end{equation}
and the leading IR divergence appears for spectral index $n\leq-1$, while 
sub-leadings divergences reveal for $n \leq -3$.  

For $Q_2(k)$, we find an IR divergence that does not appear in $\Lambda$CDM. Expanding the angular integrand of Eq.~(\ref{Q2f})
about $r=0$ (or equivalently $p=0$) with power law \typo{potential}{power spectrum} and discarding odd powers of $x$, we get
\begin{equation}
(1+r^2 - 2 r x)^{n/2}  \frac{rx(1-rx)}{1+r^2 -2 r x}
\left( \mathcal{A} - \mathcal{B} \frac{(x-r)^2}{1+r^2 -2 r x} \right) = \tocheck{(\mathcal{A} - \mathcal{B})(1-n)x^2} \frac{p^2}{k^2}  + \mathcal{O} \left( \frac{p^4}{k^4}\right)
\end{equation}
Here $\mathcal{A} = \bar{D}^{(2)}_a(0,\vk) - \bar{D}^{(2)}_{\delta I}(0,\vk)$ and  $\mathcal{B}= \bar{D}^{(2)}_{\typo{a}{b}}(0,\vk)$, 
while $\bar{D}^{(2)}_\text{FL}(0,\vk) = 0$ for this configuration. This is not the
same squeezed configuration found above in Eq.~(\ref{expintQ1}), and then $\mathcal{A} - \mathcal{B} \neq 0$. As a result we get 
\begin{equation}
 Q_2^\text{IR} = \frac{1}{6\pi^2}P_L(k)\int_0^\infty dp \, p^2 P_L(p)  \tocheck{(\mathcal{A}-\mathcal{B}) (1-n)}.
\end{equation}
Although this IR divergence does not appear in $\Lambda$CDM, it is {\it safe} since is revealed for $n\leq-3$. 
For $R_2$ and $Q_1$ functions the IR divergences when the internal momentum is sent to zero appear for $n<-3$, as in $\Lambda$CDM.

On the other hand for $\vp \rightarrow \vk$, functions $Q_1$ and $Q_2$ present divergences that goes as $(p-k)^{n+2}$ and $(p-k)^{n+1}$, respectively, 
which are of the same form as in $\Lambda$CDM.
These results rely in the approximation of letting fixed the growth functions as $\bar{D}^{(2)}(\vk,0)$ as follows from
Eq.~(\ref{Q1Q2Ev}).

For $R_2$ as $\vp \rightarrow \vk$ we have the evaluation  $\bar{D}^{(2)}(-\vk,\vk)$, which is the same squeezed 
configuration found in Eq.~(\ref{expintQ1}) for which $\mathcal{A} - \mathcal{B} = 0$. Thus, we can 
substitute $\mathcal{A}$ for $\mathcal{B}$ in the integrand of Eq.~(\ref{R2f}). Since this is a function of $k$ only it can be
factorized and pulled out of the integral. Thus $R_2$ has no IR divergence for $\vp \rightarrow \vk$ as in $\Lambda$CDM. In the absence of
frame-lagging $\mathcal{A} \neq \mathcal{B}$ and we obtain a $\log |p-k|$ divergence after expanding Eq.~(\ref{R2f}) about $p =k$ and performing the 
angular integration.  

\bigskip

The intuition developed in this section suggest that IR and UV divergences for $Q$ and $R$ functions are the same in MG and in $\Lambda$CDM.
Though, a rigorous proof is still lacking; in particular we did not analyze the divergences of $R_1$ function.
We stress that in our derivations the
role played by the frame-lagging terms is determinant, and further, we assumed MG models that converge to GR at large scales.

\bigskip

We finalize this section by showing the power spectrum $a(k) \equiv P_{\nabla \cdot \Psi \, \nabla \cdot\Psi}$ of the divergence of 
Lagrangian displacement fields, $\nabla_i \Psi^i$, at 1-loop. This is given by 
$a(k) =  k^i_1 k^j_2 \langle \Psi_i^{(1)}(k_1) \Psi_j^{(1)}(k_2) \rangle_c' 
+  k^i_1 k^j_2 \langle \Psi_i^{(2)}(k_1) \Psi_j^{(2)}(k_2) \rangle_c' + 2  k^i_1 k^j_2 \langle \Psi_i^{(1)}(k_1) \Psi_j^{(3)}(k_2) \rangle_c'$, 
where the notation $\langle (\cdots)\rangle'$ means that we omit a Dirac delta function. Using Eqs.~(\ref{defQ1}) and (\ref{defR1}) this is 
\begin{equation}
 a(k) = P_L(k) + \frac{9}{98}Q_1(k)  + \frac{10}{21} \typo{Q_2(k)}{R_1(k)}, 
\end{equation}
where $P_L(k)$ coincides with the linear power spectrum of matter fluctuations. In Fig.~\ref{fig:akplot} we show a plot of $a(k)$ with their different contributions. 
We note that the piece without frame-lagging has negative values.

\begin{figure}
	\begin{center}
	\includegraphics[width=3 in]{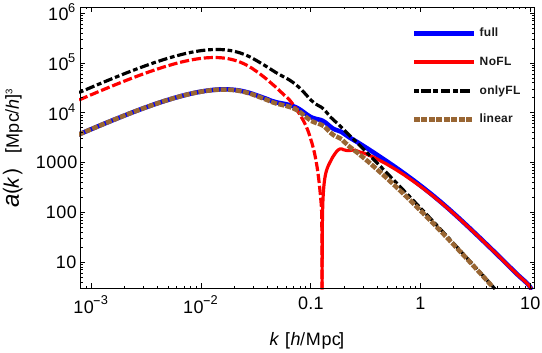}
	\caption{Power spectrum $a(k)$ of divergences of Lagrangian displacements for the F4 model: 
	The solid blue curve is the full power spectrum, the dot-dashed black contains only frame-lagging
	contributions; the red curve does not contain any frame-lagging, with the dashed showing its negative; 
	the dotted brown is for the linear theory, that coincide with the linear matter 
	power spectrum. 
	\label{fig:akplot}}
	\end{center}
\end{figure}
\end{section}

\begin{section}{CDM 2-point statistics}\label{Sect:2pStats}

In this section we compute the power spectrum and correlation function for F4, F5 and F6 models 
using the schemes described in Appendix \ref{app::correlations}. 
The building blocks are the $Q_i$ and $R_i$ functions already computed in the previous section. As discussed in the introduction,
LPT statistics are better for modeling the BAO feature while SPT is better in the broadband power spectrum, and for that reason we
explore both approaches here. Specifically, we use the Convolution Lagrangian Perturbation Theory (CLPT) of Ref.~\cite{Carlson:2012bu},
the Lagrangian Resummation Theory (LRT) \cite{Mat08a}, and the SPT* explained in Appendix \ref{app::correlations}.

For our numerical computations we fix cosmological parameters to $\Omega_m=0.281$, $\Omega_b=0.046$, $h=0.697$, 
$n_s=0.971$, and $\sigma_8=0.82$, corresponding to the best fit of the WMAP Nine-year results \cite{2013ApJS..208...19H}. The linear
$\Lambda$CDM matter power spectrum is computed with the CAMB code \cite{Lewis:1999bs}, 
and the linear MG is obtained by multiplying it by the square of linear
growth factor,
\begin{equation}
 P_L^\text{MG} (k) = D_+^2(k) P_L^\text{$\Lambda$CDM}(k).
\end{equation}

We compute the CLPT power spectrum in Eq.~(\ref{CLPTPS}) from the Hankel transformations of 
Eqs.~(\ref{PZA}), (\ref{PA}) and (\ref{PW}) and using FFTLog  numerical methods, as described 
in \cite{Hamilton:1999uv}.\footnote{In practice, the sums in Eqs.~(\ref{besselident}) are performed up to some $\ell_{max}$. In the code we use
for this work, this maximum is adaptive in the interval $k =0.001 $ -- $3 \,h/\text{Mpc}$
ranging from $\ell_\text{max} = 10$ to 30.}  
We specifically make use of the public code released in 
\cite{Vlah:2016bcl}.\footnote{\href{https://github.com/martinjameswhite/CLEFT_GSM}{https://github.com/martinjameswhite/CLEFT\_GSM.}}, that is supplied with
precomputed $\Xi_\ell$ functions, defined in \cite{Vlah:2015sea}, which are combinations of the $X$, $Y$, $V$ and $T$ 
functions defined in Eqs.~(\ref{defX})-(\ref{defTq}).

The LRT power spectrum, written in this section as $P_\text{LRT}(k) = e^{- k^2 \sigma^2_L }P_L(k) + \text{NL}$, with NL denoting the loop integrals,
and
\begin{equation}
 \sigma^2_L = \frac{1}{6\pi^2} \int_0^\infty dp P_L(p)
\end{equation}
the total variance of the divergence of linear Lagrangian displacements,
is obtained directly by providing Eq.~(\ref{MatPS}) with the $Q$ and $R$ functions. 

Since LRT and CLPT power spectra decay quickly (in $k$), we expand them to obtain 
the 1-loop SPT* power spectrum ---which in EdS coincides with the 1-loop SPT power spectrum--- as explained in 
the Appendix \ref{app::correlations}. Thus, we analyze also the power spectrum  
$P_{\text{SPT}^*}(k) = (1-\sigma^2_L k^2)P_L(k) + \text{NL}$ of Eq.~(\ref{PSPTstar}).

\begin{figure}
	\begin{center}
	\includegraphics[width=3 in]{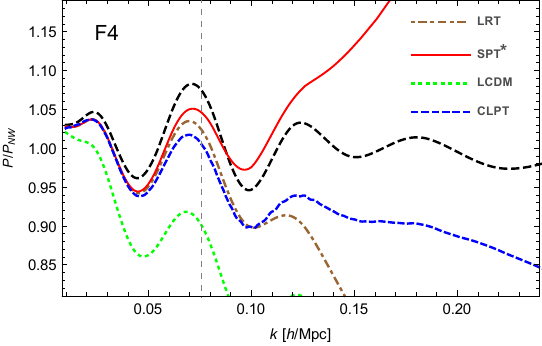}
	\hspace{0.3cm}
	\includegraphics[width=3 in]{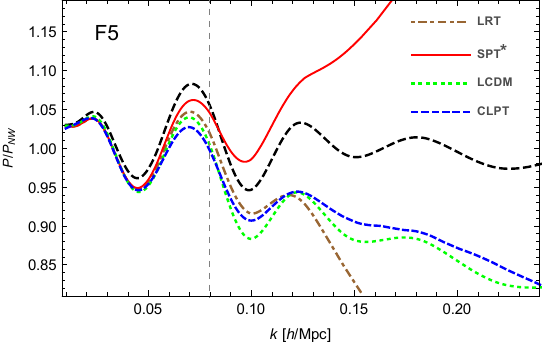}
	\caption{Ratios of power spectra to the linear corresponding models (F4 and F5) 
	power spectrum with BAO removed at redshift
	$z=0$. Long dashed black curves are for the linear
	theory; dashed blue are for CLPT; solid red are for SPT*; dot-dashed brown are for LRT; 
	and dotted green are the linear $\Lambda$CDM. 
	The vertical line shows the scale $k_\text{NL}=\sigma^{-1}_{L}/2$.
	\label{fig:PSoverNWF4F5}}
	\end{center}
\end{figure}

\begin{figure}
	\begin{center}
	\includegraphics[width=3 in]{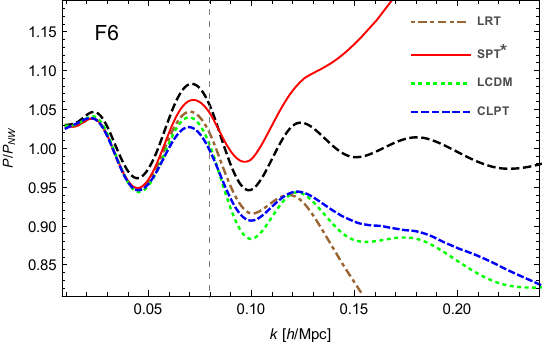}
	\hspace{0.3cm}
	\includegraphics[width=3 in]{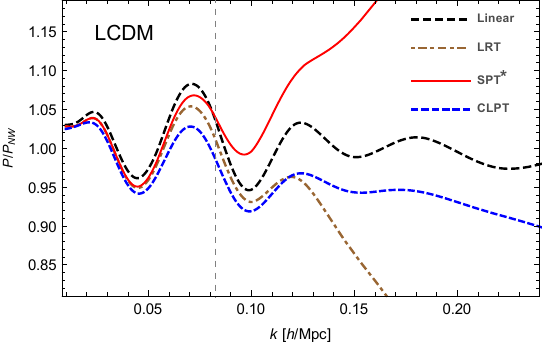}
	\caption{Ratios of power spectra to the linear corresponding models (F6 and $\Lambda$CDM) 
	power spectrum with BAO removed at redshift
	$z=0$. Long dashed black curves are for the linear
	theory; dashed blue are for CLPT; solid red are for SPT*; dot-dashed brown are for LRT.
	The F6 panel shows also the linear $\Lambda$CDM case, denoted with dotted green curve. 
	The vertical line shows the scale $k_\text{NL}=\sigma^{-1}_{L}/2$.
	\label{fig:PSoverNWF6LCDM}}
	\end{center}
\end{figure}

In Figs.~\ref{fig:PSoverNWF4F5} and \ref{fig:PSoverNWF6LCDM} we show the ratio of the different power spectra and the linear 
power spectrum of the considered model with the BAO removed, this is done for
CLPT, LRT and SPT*, and for F4, F5, F6 and $\Lambda$CDM models at redshift $z=0$. 
Dashed blue curves correspond to the CLPT scheme, solid red are for the SPT*, brown dashed for LRT, and long-dashed black for
the linear theory. As a reference, for MG models we plot the linear $\Lambda$CDM model as well (depicted by dotted green curves).
The vertical lines 
denote the expected maximum \typo{wavelengths}{wavenumbers} $k_{max}$ of validity of the perturbative Lagrangian formalisms. For these, we adopt the prescription in
Ref.~\cite{Mat08a} relying on the damping scale, such that $k_{max}=\sigma_\text{L}^{-1}/2$; other prescriptions are possible, but do not differ substantially. 
We note that in MG models non-linearities starts to be
more relevant at larger scales since clustering is more efficient due to the extra force.

\begin{figure}
	\begin{center}
	\includegraphics[width=3 in]{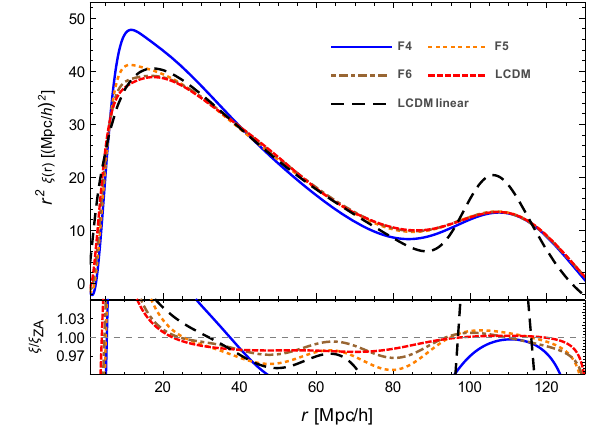}
	\caption{CLPT correlation function at redshift $z=0$. Long dashed (black) is the linear
	$\Lambda$CDM model; dashed (red) is for $\Lambda$CDM in CLPT ; solid (blue) is for F4; dotted (orange) for F5; 
	and dot-dashed (brown) is for F6. The lower panel shows the ratio with the Zel'dovich approximation in $\Lambda$CDM.
	\label{fig:CF}}
	\end{center}
\end{figure}

LPT formulations perform better for the correlation function; it was indeed noted that even the Zel'dovich approximation
provides very good accuracy about the BAO peak; see for example \cite{Tassev:2013rta,Whi14} for recent discussions. We compute the 
CLPT correlation functions given by Eq.~(\ref{CLPTcf}). To do it we feed the code 
of \cite{Vlah:2015sea}\footnote{\href{https://github.com/alejandroaviles/CLEFT}{https://github.com/alejandroaviles/CLEFT.}} with
precomputed  $Q_i$ and $R_i$ functions. 
We show the results in Fig.~\ref{fig:CF}, with the lower
panel showing the relative difference when comparing to the $\Lambda$CDM Zel'dovich approximation in the $\Lambda$CDM,  
which is computed from Eq.~(\ref{PZA}).
At the BAO scale, the differences in the correlation among the three
MG and $\Lambda$CDM models are smaller than $2\%$ 
since at these scales the fifth force is almost negligible. Soon below the BAO scales the differences 
depart considerably for F4, which is not surprising since this model is ruled out even at the linear level \cite{Hojjati:2015ojt}.

In Figs.~\ref{fig:PSF4F5} and \ref{fig:PSF6} we show the SPT* power spectra for two redshifts $z=0$ and $z=0.5$ with and without
screenings.  As a reference we use  
$N$-body simulations data \cite{Liaknowled} performed with the ECOSMOG code of Ref.~\cite{Li:2011}. The presented data
are the average of two realizations in boxes of size $1024\,\text{Mpc}/h$ and  with $1024^3$ dark matter particles. 
It is difficult to tell what is the scale at which SPT* is valid since the data is noisy at large 
scales,\footnote{We stress that $N$-body codes for MG are numerically quite expensive and are aimed mainly to test the small scales. Thus 
they use relatively small boxes and large scales turn out to be highly affected by cosmic variance.}  
but we expect to have a good
accuracy up to the $k_{max}$ given above. Nevertheless, we can note that the trend line of the data is properly followed by the analytical
method.

\begin{figure}
	\begin{center}
	\includegraphics[width=3 in]{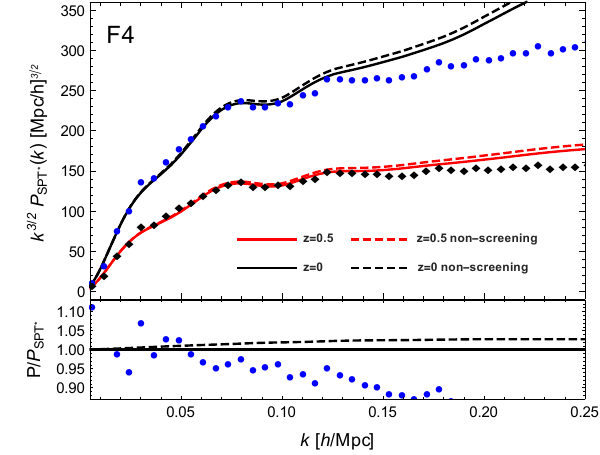}
	\includegraphics[width=3 in]{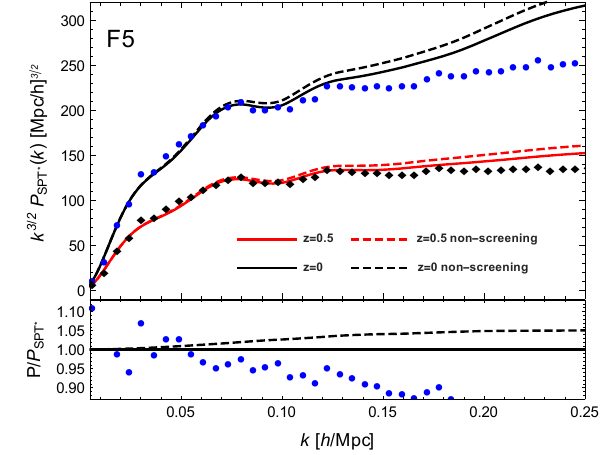}
	\caption{F4 and F5. SPT* power spectra with (solid curves) and without (dashed curves) screenings for models F4 and F5, at redshifts 
	$z=0$ (black curves) and $z=0.5$ (red curves). The lower panels shows the ratio with the SPT* power spectra for $z=0$. The simulated
	data were provided by \cite{Liaknowled}.
	\label{fig:PSF4F5}}
	\end{center}
\end{figure}

\begin{figure}
	\begin{center}
	\includegraphics[width=3 in]{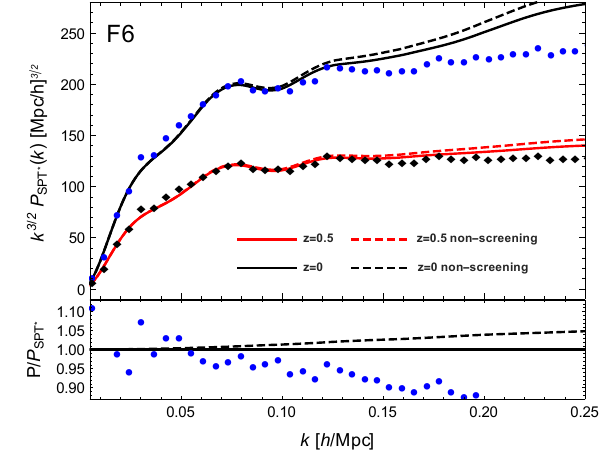}
	\caption{Same as Fig.~\ref{fig:PSF4F5} but for F6.
	\label{fig:PSF6}}
	\end{center}
\end{figure} 

\bigskip

Since small scales structures are subject to negligible tidal forces from bulk flows provided by long wavelength fluctuations, these are only affected by
an overall translation which has no net effect in their matter $N$-point statistics. The flaw of LRT is that bulk flows are split in $q$ dependent and independent
pieces [see Eq.~(\ref{LRTsplit})] and does not cancel at zero Lagrangian separation. Indeed, 
in the exponential prefactor of Eq.~(\ref{MatPS}) we find the term  $\sigma^2_L$, 
that has an IR divergence for power law power spectrum with spectral index $n\leq-1$, as a consequence  
the bulk flows largely affect the small scales statistics, this is a manifestation of the breaking the Galilean invariance (see for example 
\cite{Sugiyama:2013mpa} for recent discussion in LRT). This  effect 
is observed in Fig.~\ref{fig:PSoverNWF4F5} and \ref{fig:PSoverNWF6LCDM} where a deficit in the 
power spectrum with respect to the CLPT results is quite evident. Instead, in CLPT all
linear contributions are kept in the exponential [see Eq.~(\ref{CLPTPS})], including the term  
\begin{equation}
\frac{1}{2}X_L(q) = \frac{1}{2\pi^2} \int dk \,P(k)\left( \frac{1}{3} - \frac{j_1(k q)}{k q}\right) 
\, \underset{k\rightarrow 0}{\longrightarrow}  \, \frac{q^2}{60 \pi^2}  \int dk \, k^2 P(k), 
\end{equation}
which is IR-safe for $n > -3$.\footnote{For power spectrum with $n\leq -3$, tidal forces from long modes can not be
neglected at any scale \cite{Carrasco:2013sva}.}

The SPT* power spectrum is usually written as  $P_\text{SPT*}(k) = P_L(k) + P_{22}(k) + P_{13}(k)$, in terms of $Q$ and $R$ functions these are
\begin{align}
P_{22}(k) &= \frac{9}{98}Q_1(k) + \frac{3}{7}Q_2(k) + \frac{1}{2} Q_3(k), \label{P22}\\
P_{\typo{11}{13}}(k) &=  \frac{10}{21}R_1(k) + \frac{6}{7}R_2(k) - \sigma_L^2 k^2 P_L(k). \label{P13}
\end{align}
The loop integrals have in general IR divergences for power law power spectrum with spectral index $n<-1$, 
but they cancel out exactly when the sum is considered. 
This is a consequence of the discussion above but is more general due to the equivalence 
principle \cite{Carrasco:2013sva,Creminelli:2013mca}.
We have already found in Eq.~(\ref{Q3IR})  IR divergences in $Q_3$ when $p\rightarrow 0$ and $p \sim k$, which are twice that of $\sigma^2_L P_L$,
but in Eq.~(\ref{P22}) $Q_3$ is multiplied by a $1/2$ factor; thus, when summing up the two SPT leading divergences cancel. This is an old 
known result since the work of \cite{1983MNRAS.203..345V}, and afterwards noted in \cite{1996ApJ...456...43J,Scoccimarro:1995if} 
that the same should happen for sub-leading divergences; we indeed have found other sub-leading divergences for $R$ and $Q$ functions.  
Our reasoning of Sect.~\ref{Sect:LDstats} suggests that the additional, sub-leading
IR divergences with power spectral index $n>-3$ are the same as those of $\Lambda$CDM, making the theory IR-safe. 

Even for MG theories with universal couplings, the principle of equivalence can be violated since some bodies 
may develop screening against the additional force while some others may not \cite{Hui:2009kc}, 
but for long wavelength fluctuations this is irrelevant as long as they essentially exert only the Newtonian force upon the small structures.
A brute force approach is elusive, but the absence of IR divergences (for $n>-3$) in MG theories that reduce to GR at large scales 
is in principle ensured by the equivalence principle \cite{Valageas:2013cma}.

\end{section}

\begin{section}{Conclusions} \label{Sec:concl}

In this work we found a generalization to alternative theories of gravity for the Lagrangian displacement field up to third order in perturbation theory.
In doing this, an LPT theory for the study of large scale structure 
formation  was developed. 
Our formalism is suitable for Horndeski models \cite{Bose:2016qun}, this is the same range of validity that
share several investigations in SPT that came after to the pioneering work of Ref.~\cite{2009PhRvD..79l3512K}. 
The basic requirement is that the theory can be \typo{recasted}{recast} as
a scalar tensor gravity at linear order, all non linear contributions to the Klein-Gordon equation, including possible 
kinetic terms, can be written as sources and treated perturbatively. 
In theories that have a screening mechanism these contributions are responsible to the recover GR at small scales, 
by making the Laplacian of the scalar field negligible.

Since in LPT is standard to work in $q$-Fourier space, one needs to transform the Poisson equation accordingly. In these theories the Poisson equation has two
sources: the standard source of matter perturbations can be written in terms of Lagrangian displacement straightforwardly due to mass conservation, 
but the terms arising from the scalar field must be properly transformed. By doing it, terms to compensate spatial derivatives in Eulerian space appear
into the theory, we call these geometrical contributions frame-lagging terms, and have a crucial role in the LPT for MG, which is more evident at large scales 
where the theory should reduce to GR. We have discussed this limit from different perspectives. A special one is the 1-dimensional collapse for which 
the Zel'dovich approximation is exact (up to shell-crossing) in $\Lambda$CDM, while in MG this is not the case since the force mediated by the 
scalar field does not decay as the square of the inverse of the distance. But at scales larger than the range of the fifth force the only force in play is
the standard Newtonian force and  
we have to recover Zel'dovich as the exact solution, or in other words, the Lagrangian displacement growth functions at orders higher than 1 should vanish. 
The formalism presented here is capable to capture this physical fact, mathematically through cancellations provided by the frame-lagging contributions. 

Throughout this article we apply the formalism by using the Hu-\typo{Sawicky}{Sawicki} $f(R)$ theory \cite{Hu:2007nk}, more precisely to the so-called F4, F5, and F6 models.
We compute 2-point statistics both for the Lagrangian displacement and for the matter perturbations. For the matter power spectrum 
we use recent $N$-body simulations \cite{Liaknowled} as a reference. 
LPT is very successful in reproducing the acoustic peak in the correlation function, and we use two different resummation schemes: CLPT and LRT. 
Though, LPT is not satisfactory in following the broadband
power spectrum. For this reason, we also use a scheme that reduces to SPT for the EdS case  \cite{Mat08a} ---in MG it is not clear if one 
recovers SPT exactly since frame-lagging terms do not seem to cancel out--- showing good agreement between the analytic theory and simulations.

We further discussed UV and IR divergences in matter and Lagrangian displacement statistics. Our analysis is not 
exhaustive, but it suggests that the schemes of SPT* and CLPT are IR-safe in MG.

\end{section}

\begin{acknowledgments}
The authors would like to thank Baojiu Li for useful discussions and for sharing the $N$-body simulation data. We also thank to 
Zvonimir Vlah and Matteo Cataneo for suggestions and discussions. 
A.A. is supported by
C\'atedras CONACyT project No.574. The authors acknowledge financial support from CONACyT Project
269652 and Fronteras Project 281. 
\end{acknowledgments}

\appendix

\begin{section}{Lagrangian displacement polyspectra and the matter power spectrum} \label{app::correlations}

This appendix briefly reviews the methods we use to compute the power spectrum and the correlation function. For completeness
we write the relevant equations but we refer the reader to the original papers for detailed derivations.

The matter power spectrum in LPT is given by \cite{TayHam96}
\begin{equation}\label{app2:PLPT}
 P_{\text{LPT}}(\vk) = \int d^3 q e^{-i\vk \cdot \vq} \left( \langle e^{-i\vk \cdot \Delta}\rangle - 1 \right),
\end{equation}
where $\Delta_i = \Psi_i(\vq_2) - \Psi_i (\vq_1)$ are the Lagrangian displacement differences, and $\vq = \vq_2-\vq_1$ 
the Lagrangian coordinate difference.
By using the cumulant expansion theorem  we can recast the power spectrum as \cite{Carlson:2012bu}
\begin{equation}\label{app2:PLPT1loop}
 \typo{(2\pi)^2}{(2\pi)^3}\delta_D(\vk) +  P_{\text{LPT}}(\vk) = \int d^3 q  e^{-i\vk \cdot \vq}  \exp \left[- \frac{1}{2} k_i k_j A_{ij}(\vq) 
  + \frac{i}{6}k_i k_j k _k W_{ijk}(\vq) \right],
\end{equation}
where $A_{ij}=\langle \Delta_i \Delta_j \rangle_c$ and $W_{ijk}=\langle \Delta_i \Delta_j \Delta_k \rangle_c$.
Here we show terms up to third order in the Lagrangian displacement since we are interested in the first, 
1-loop corrections to the matter two point functions. 
The products  $\Delta_i\Delta_j$ have contributions evaluated at the same point, the so-called zero-lag, and contributions given at a separation $\vq$. 

The direct computation of the LPT power spectrum of Eq.~(\ref{app2:PLPT1loop}) 
is difficult since it involves highly oscillatory integrands. This has been done in 
\cite{Vlah:2014nta} by using expansions in spherical Bessel functions (see below) 
and in \cite{Sugiyama:2013mpa} by similar methods involving Legendre expansions of the Lagrangian displacement differences. 
In part because of these numerical complications, alternative resummation schemes that Taylor expand some terms in the 
exponential of Eq.~(\ref{app2:PLPT1loop}) 
exist in the literature. We below review those we use in this work.

We first define the polyspectra $C^{(n_1 \cdots n_N)}_{i_1\cdots i_N}(\vk_1,\dots,\vk_N)$ at order $(n_1 +\cdots + n_N)$ as in \cite{Mat08a}
\begin{equation} \label{MatPol}
 \langle \Psi^{(n_1)i_1}(\vk_1)\cdots\Psi^{(n_N) i_N}(\vk_N)\rangle_c = 
 (-i)^{N-2} (2\pi)^3 \delta_D (\vk_1 + \cdots + \vk_N) C^{(n_1 \cdots n_N)}_{i_1\cdots i_N}(\vk_1,\dots,\vk_N),
\end{equation}
which in terms of the Lagrangian displacement kernels can be written as
\begin{align}
 C^{(11)}_{ij}(\vk) &= L^{(1)}_i(\vk) L^{(1)}_j(\vk) P_L(k), \label{c11}\\
  C^{(22)}_{ij}(\vk) &=  \frac{1}{2}\int \frac{d^3p}{(2 \pi)^3} L^{(2)}_i(\vp,\vk-\vp)L^{(2)}_j(\vp,\vk-\vp) 
P_L(p) P_L(|\vk-\vp|),   \label{C22} \\
  C^{(13)}_{ij}(\vk) =C^{(31)}_{ij}(\vk)&= 
    \frac{1}{2}L^{(1)}_i(\vk) P_L(k) \int \frac{d^3p}{(2 \pi)^3} L^{(3)symm}_j(\vk,-\vp,\vp) P_L(p),  \label{c13} \\
C^{(112)}_{ijk}(\vk_1,\vk_2,\vk_3) =  C^{(121)}_{jki}(\vk_2,\vk_3,\vk_1) &= C^{(211)}_{kij}(\vk_3,\vk_1,\typo{\vk_1}{\vk_2})
= -L^{(1)}_i(\vk_1) L^{(1)}_j(\vk_2) L^{(2)}_k(\vk_1,\vk_2) P_L(k_1) P_L(k_2). \label{c112}
\end{align}
Higher order polyspectra do not enter in 1-loop calculations. The following scalar functions are constructed
\begin{align}
 Q_1(k) &= \frac{98}{9} k_i k_j C_{ij}^{(22)}(k), \label{defQ1}\\
 Q_2(k) &= \frac{7}{3} k_i k_j k_k \int \Dk{p} C_{ijk}^{(211)}(\vk,-\vp,\vp-\vk), \label{defQ2}\\
 Q_3(k) &= k_i k_j k_k k_l  \int \Dk{p}  C^{(11)}_{ij}(\vp)C^{(11)}_{kl}(\vk - \vp), \label{defQ3}\\
 R_1(k) &= \frac{21}{5} k_i k_j C_{ij}^{(13)}(k), \label{defR1}\\
 R_2(k) &= \frac{7}{3} k_i k_j k_k \int \Dk{p} C_{ijk}^{(112)}(\vk,-\vp,\vp-\vk) = \frac{7}{3} k_i k_j k_k \int \Dk{p} C_{ijk}^{(121)}(\vk,-\vp,\vp-\vk).\label{defR2}
\end{align}
Notice these functions are all of order $\mathcal{O}(P_L^2)$.
By using decomposition in vectors $\hat{q}_i$ the spectra and bispectra of Lagrangian displacement 
differences take the form \cite{Carlson:2012bu,Vlah:2014nta}
\begin{align}
 A_{ij}(\vq) &= X(q)\delta_{ij} + Y(q) \hat{q}_i \hat{q}_j \\
 W_{ijk}(\vq) &= V(q) \hat{q}_{\{i}\delta_{jk\}} + T(q) \hat{q}_i \hat{q}_j\hat{q}_k
\end{align}
with
\begin{align}
 X(q) &= \frac{1}{\pi^2} \int^\infty_0 dk \left( P_L (k) + \frac{9}{98} Q_1(k) + \frac{10}{21} R_1(k)\right) \left(\frac{1}{3} - \frac{j_1(k q)}{k q} \right) \label{defX} \\
 Y(q) &= \frac{1}{\pi^2} \int^\infty_0 dk \left( P_L (k) + \frac{9}{98} Q_1(k) + \frac{10}{21} R_1(k)\right) j_2(k q) \label{defY}\\
 V(q) &= -\frac{3}{21 \pi^2} \int_0^\infty \frac{dk}{k}(Q_1(k) - 3Q_2(k) + 2 R_1(k) - 6 R_2(k)) j_1(\typo{k_q}{kq}) - \frac{1}{5} T(q), \label{defV}\\
 T(q) &=   -\frac{9}{14 \pi^2} \int_0^\infty \frac{dk}{k}(Q_1(k) + 2 Q_2(k) + 2 R_1(k) + 4 R_2(k)) j_3(\typo{k_q}{kq}), \label{defTq}
\end{align}
where $ P_L (k)$ is the linear matter power spectrum. (In general ``$L$'' denotes a linear piece and ``loop'' a pure 1-loop piece.) 
These functions vanish as Lagrangian coordinates separation goes to zero, implying that the power spectrum does not have contribution
for arbitrary small scales. This is consistent with Eq.~(\ref{app2:PLPT}).

The formalism of Convolution Lagrangian Perturbation Theory (CLPT) as presented in \cite{Vlah:2015sea} 
relies on expanding the loop contributions of Eq.~(\ref{app2:PLPT1loop}) while keeping the linear terms in the 
exponential,\footnote{CLPT was introduced in \cite{Carlson:2012bu} where also the $A_{ij}^{\text{loop}}$ was kept exponentiated.}
leading to 
\begin{equation} \label{CLPTPS}
(2\pi)^3 \delta_\text{D}(\vk) + P_{\text{CLPT}}(k) = \int d^3 q e^{-i\vk \cdot \vq} \exp \left( -\frac{1}{2}k_{i}k_{j}A_{ij}^{L}(\vq) \right) 
\left[1 -\frac{1}{2}k_{i}k_{j}A_{ij}^{\text{loop}}(\vq) + \frac{i}{6} k^{i}k^{j}k^k W_{ijk}^{\text{loop}}(\vq) \right].
\end{equation}
This can be separated in Zel'dovich and loops contributions as 
\begin{equation}
 P_{\text{CLPT}}(k) = P_\text{ZA}(k) + P_{A}(k) + P_{W}(k)
\end{equation}
with (for $\vk \neq 0$)
\begin{align}
  P_\text{ZA}(k) &=  2 \pi \int_0^{\infty} d q \, q^2 e^{- \frac{1}{2} k^2 X_{L}} 
  \int^1_{-1} d \mu \exp \left(i \mu k q  - \frac{1}{2} \mu^2 k^2 Y_L \right)  \label{PZA}\\
  P_{A}(k)  &= - 2 \pi \int_0^{\infty} d q \, q^2 e^{- \frac{1}{2} k^2 X_L } 
                  \int^1_{-1} d \mu \left(   \frac{1}{2} k^2 X_{\text{loop}} +  \frac{1}{2} k^2 \mu^2  Y_{\text{loop}} \right)
                    \exp \left(i \mu k q  - \frac{1}{2} \mu^2 k^2 Y_L \right) \label{PA} \\
  P_W(k) &= - 2 \pi \int_0^{\infty} d q \, q^2 e^{- \frac{1}{2} k^2 X_L} 
  \int^1_{-1} d \mu \left(   \frac{i}{2} k^3 \mu V  +  \typo{\frac{i}{2} k^3 \mu^3  T   \frac{1}{2} k^2 Y_{\text{loop}}}{\frac{i}{6} k^3 \mu^3  T} \right) 
  \exp \left(i \mu k q  - \frac{1}{2} \mu^2 k^2 Y_L \right).  \label{PW}
 \end{align}
By using the expansions \cite{Vlah:2014nta,Vlah:2015sea}
\begin{align} 
  \int^1_{-1} d \mu e^{i \mu A + \mu^2 B} &= 2 e^B \sum_{\ell=0}^{\infty} \left(-\frac{2 B}{A} \right)^{\ell} j_{\ell}(A), \\
  \int^1_{-1} d \mu \mu e^{i \mu A + \mu^2 B} &= 2 i e^B \sum_{\ell=0}^{\infty} \left(-\frac{2 B}{A} \right)^{\ell} j_{\ell+1}(A), \\
  \int^1_{-1} d \mu \mu^2 e^{i \mu A + \mu^2 B} &= 2 e^B \sum_{\ell=0}^{\infty} \left(1+\frac{\typo{l}{\ell}}{B} \right)\left(-\frac{2 B}{A} \right)^{\ell} j_{\ell}(A), \\
  \int^1_{-1} d \mu \mu^3 e^{i \mu A + \mu^2 B} &= 2 i e^B \sum_{\ell=0}^{\infty} \left(1+\frac{\typo{l}{\ell}}{B} \right)\left(-\frac{2 B}{A} \right)^{\ell} j_{\ell+1}(A), \label{besselident}
\end{align}
the integrals in  Eqs.(\ref{PZA}), (\ref{PA}) and (\ref{PW}) reduce to one-dimensional Hankel 
transformations.

In the CLPT scheme, the correlation function can be written compactly by Fourier transforming the power spectrum of Eq.~(\ref{CLPTPS}) and using Gaussian integrations, resulting in
\begin{equation} \label{CLPTcf}
 1+ \xi_{\text{CLPT}}(r) = \int \frac{d^3 q}{ \typo{(2 \pi)^3 \det[A^L_{ij}]^2}{(2 \pi)^{3/2} \det[A^L_{ij}]^{1/2}}} 
 e^{\typo{\frac{1}{2} A^L_{ij}}{-\frac{1}{2} (A^L)^{-1}_{ij}} (r_i - q_i)(r_j - q_j)} \left[ 1-\frac{1}{2}G_{ij} A^{\text{loop}}_{ij} 
 + \frac{1}{6} \Gamma_{ijk} W_{ijk}^{\text{loop}} \right]
\end{equation}
where $G_{ij} = A^{-1}_{ij} - g_ig_j$, $\Gamma_{ijk} = A^{-1}_{ij} g_k +A^{-1}_{jk} g_i + A^{-1}_{ki} g_j - g_ig_jg_k$, 
and $g_i = A^{-1}_{ij}(r_j - q_j)$. 

\bigskip

Now, a straightforward expansion of the exponential in Eq.~(\ref{CLPTPS}) leads to \cite{Vlah:2014nta,Vlah:2015sea}
\begin{equation} \label{PSPTstar}
 P_{\text{SPT*}} =  P_L(k) 
 + \frac{10}{21}R_1(k) + \frac{6}{7} R_2(k)
 + \frac{9}{98}Q_1(k)  + \frac{3}{7} Q_2(k) + \frac{1}{2} Q_3(k) - \sigma^2_L k^2 P_L(k)
\end{equation}
where
\begin{equation}
 \sigma^2_L = \frac{1}{2} X_L(q\rightarrow \infty) =  \frac{1}{6 \pi^2} \int^\infty_0 dk  P_L (k)
\end{equation}
is the 1-dimensional Lagrangian displacement variance. The power spectrum of Eq.~(\ref{PSPTstar}) 
coincides with the SPT power spectrum in the \lcdm case. Note that this may not be
the case for MG, since it is not clear that the frame-lagging terms cancel out, and SPT should be free of these terms. Nevertheless, the behavior the 
power spectrum in this scheme is similar to what is expected from an SPT theory, 
then we use it in Sect.~\ref{Sect:2pStats} and denote it as SPT*.

\vspace{0.2cm}

Other scheme widely used in the literature is the Lagrangian Resummation Theory (LRT) of Matsubara \cite{Mat08a}, here the $A_{ij}$ matrix splits as
\begin{equation}\label{LRTsplit}
 A_{ij} = X(q \rightarrow \infty) \delta_{ij} + (X(q) - X(q\rightarrow \infty)) \delta_{ij} + Y(q) \hat{q}_i \hat{q}_j 
\end{equation}
and the zero-lag term $X(q \rightarrow \infty)=2\sigma^2$ is kept in the exponential while the rest is expanded, leading to 
\begin{equation} \label{MatPS}
 P_\text{1-loop}^{\text{LRT}}(\vk) = e^{ -k^2 \sigma^2_L }   
 \left( P_L(k) + \frac{9}{98}Q_1(k)  + \frac{10}{21}R_1(k) + \frac{3}{7} Q_2(k) + \frac{6}{7} R_2(k) + \frac{1}{2} Q_3(k) \right).
\end{equation}
(This equation coincides with Eq.~(35) of \cite{Mat08a}.) By a further expansion of the exponential we arrive back at Eq.~(\ref{PSPTstar}).

The exponential prefactor in Eq.~(\ref{MatPS}) is responsible for the smearing of the BAO peak, and leads to the suppression
of the power spectrum at small scales. Note also that the loop variance $\sigma^2_\text{loop}$  
is neglected in the exponential since it is already a $\mathcal{O}(P_L^2)$ quantity. If it is considered, the corrections become large and the BAO peak is over-suppressed 
\cite{Vlah:2014nta}. 

We note that LRT is inconsistent since $A_{ij}$ should vanish at small separations.
This is a manifestation of the breaking of Galilean invariance in the Matsubara formalism, and it is discussed in Sect.\ref{Sect:2pStats}. 
Nevertheless, this scheme has the advantages to be computationally simpler, 
it is straightforward to add biased tracers and RSD \cite{Matsubara:2008wx}, 
and furthermore, its correlation function shows very good agreements with $N$-body simulations. 
For these reasons, we also consider it in this work.

\end{section}

\begin{section}{Computation of growth functions}\label{app::kernels}

\begin{subsection}{Second order}\label{app::kernels2}
To second order, the equation of motion (Eq.~(\ref{eqm})) can be written as
\begin{align}\label{eqm2}
 (\T - A(k)) [\Psi_{i,i}^{(2)}](\vk) &= [\Psi_{i,j}^{(1)} \T \Psi_{j,i}^{(1)}](\vk) 
 - \frac{A(k)}{2}[\Psi_{i,i}^{(1)}\Psi_{j,j}^{(1)} + \Psi_{i,j}^{(1)} \Psi_{j,i}^{(1)}](\vk)
  + \frac{k^2/a^2}{6 \Pi(\vk)} \delta I^{(2)} (\vk)  \nonumber\\
 &\quad+ \frac{M_1}{3 \Pi(k)}  \frac{1}{2 a^2}[(\nabla^2_\vx \varphi - \nabla^2 \varphi)]^{(2)}(\vk) 
 \equiv S_1 + S_2 + S_3 + S_4.
\end{align} 
\end{subsection}
We compute the result for each source $S_i$:

\bigskip

\underline{Source $S_1$:}
\begin{align}
 (\T - A(k)) [\Psi_{S_1 \, i,i}^{(2)}](\vk) &= [\Psi_{i,j}^{(1)} \T \Psi_{j,i}^{(1)}](\vk)
 = \ikk [\Psi_{i,j}^{(1)}](\vk_1) \T  [\Psi_{i,j}^{(1)}](\vk_2) \nonumber\\
 &= \ikk \frac{(\vk_1 \cdot \vk_2)^2}{k^2_1 k^2_2 } D_+(\vk_1) \T D_+(\vk_2) \delta_1 \delta_2 
 = \ikk \frac{(\vk_1 \cdot \vk_2)^2}{k^2_1 k^2_2 } A(k_2) D_+(\vk_1) D_+(\vk_2) \delta_1 \delta_2 \nonumber\\
 &= \frac{1}{2} \ikk \frac{(\vk_1 \cdot \vk_2)^2}{k^2_1 k^2_2 } (A(k_1) + A(k_2)) D_+(\vk_1) D_+(\vk_2) \delta_1 \delta_2.
\end{align}
In the fourth equality we used the linear order equation $\T D_+(k_2) = A(k_2) D_+(k_2)$, and in the fifth equality we
symmetrized the integrand. We get for the displacement field
\begin{equation} \label{ld2S1}
 \Psi_{S_1}^{(2)i} = -\frac{i}{2} \frac{k^i}{k^2} \ikk D^{(2)}_{S_1}(\vk_1,\vk_2)\frac{(\vk_1 \cdot \vk_2)^2}{k^2_1 k^2_2 }  \delta_1 \delta_2
\end{equation}
with 
\begin{equation}\label{ld2S1d}
  D^{(2)}_{S_1}(\vk_1,\vk_2) = (\T - A(k))^{-1}\big( (A(k_1) + A(k_2)) D_+(\vk_1) D_+(\vk_2)  \big)
\end{equation}

\underline{Source $S_2$:}
\begin{align}
 (\T - A(k)) [\Psi_{S_2 \, i,i}^{(2)}](\vk) &= -\frac{A(k)}{2}[\Psi_{i,i}^{(1)} \Psi_{j,j}^{(1)} + \Psi_{i,j}^{(1)} \Psi_{j,i}^{(1)}](\vk)
 = -\frac{A(k)}{2} \ikk \left(1 + \frac{(\vk_1 \cdot \vk_2)^2}{k^2_1 k^2_2 }\right) D_+(\vk_1)  D_+(\vk_2) \delta_1 \delta_2  \nonumber\\
 &= -\frac{1}{2}\ikk \left(1 + \frac{(\vk_1 \cdot \vk_2)^2}{k^2_1 k^2_2 }\right) A(k) D_+(\vk_1) D_+(\vk_2) \delta_1 \delta_2, 
\end{align}
and the displacement field becomes
\begin{equation} \label{ld2S2}
 \Psi_{S_2}^{(2)i} = \frac{i}{2}  \frac{k^i}{k^2}  \ikk D^{(2)}_{S_2}(\vk_1,\vk_2)\left(1 + \frac{(\vk_1 \cdot \vk_2)^2}{k^2_1 k^2_2 }\right) \delta_1 \delta_2,
\end{equation}
with 
\begin{equation} \label{ld2S2d}
  D^{(2)}_{S_2}(\vk_1,\vk_2) = (\T - A(k))^{-1}\big( A(k) D_+(\vk_1) D_+(\vk_2)  \big).
\end{equation}

Source $S_3$ corresponds to the screening and it leads to 
\begin{equation} \label{ld2S3}
 \Psi_{\delta I}^{(2)i} = -\frac{i}{2} \frac{k^i}{k^2}  \ikk D^{(2)}_{\delta I}(\vk_1,\vk_2) \delta_1 \delta_2
\end{equation}
with 
\begin{equation} \label{ld2S3d}
 D^{(2)}_{\delta I}(\vk_1,\vk_2) = \big(\T - A(k)\big)^{-1} 
\left( \left(\frac{2 A_0}{3}\right)^2 \frac{k^2}{a^2}\frac{M_2(\vk_1,\vk_2) D_+(k_1) D_+(k_2)}{6 \Pi(k)\Pi(k_1)\Pi(k_2)} \right) 
\end{equation}

Source $S_4$ corresponds to the frame-lagging source, leading to 
\begin{equation} \label{ld2S4}
 \Psi_\text{FL}^{(2)i} = \frac{i}{2}  \frac{k^i}{k^2}  \ikk D^{(2)}_\text{FL}(\vk_1,\vk_2) \delta_1 \delta_2
\end{equation}
with 
\begin{equation} \label{ld2S4d}
D^{(2)}_\text{FL}(\vk_1,\vk_2) =  \big(\T - A(k)\big)^{-1}  \left(  \frac{M_1(k)}{3 \Pi(k)} 
 \mathcal{K}^{(2)}_\text{FL}(\vk_1,\vk_2)   D_+(k_1) D_+(k_2) \right) 
\end{equation}
Using Eqs.~(\ref{ld2S1}),(\ref{ld2S1d}),(\ref{ld2S2}),(\ref{ld2S2d}),(\ref{ld2S3}),(\ref{ld2S3d}),(\ref{ld2S4}), and (\ref{ld2S4d}),
and rearranging terms we arrive to Eq.~(\ref{LD2order}).

\begin{subsection}{Third order}\label{app::kernels3}
 
 The equation of motion to third order is given by
\begin{align} \label{eqm3}
 (\T - A(k)) [\Psi_{i,i}^{(3)}](\vk) &= [\Psi_{i,j}^{(2)} \T \Psi_{j,i}^{(1)}](\vk) +[\Psi_{i,j}^{(1)} \T \Psi_{j,i}^{(2)}](\vk) 
 - A(k)[\Psi_{i,j}^{(2)} \Psi_{j,i}^{(1)}](\vk) - A(k) [\Psi_{i,i}^{(1)}\Psi_{j,j}^{(2)}](\vk)  \nonumber\\
 & - [\Psi_{i,k}^{(1)}\Psi_{k,j}^{(1)} \T \Psi_{j,i}^{(1)}](\vk) + \frac{A(k)}{3} [\Psi_{i,k}^{(1)}\Psi_{k,j}^{(1)} \Psi_{j,i}^{(1)}](\vk) 
 +\frac{A(k)}{6} [\Psi_{i,i}^{(1)}\Psi_{j,j}^{(1)}\Psi_{k,k}^{(1)}](\vk)  \nonumber\\
  &  + \frac{A(k)}{2} [\Psi_{l,l}^{(1)} \Psi_{i,j}^{(1)} \Psi_{j,i}^{(1)}](\vk) + \frac{k^2/a^2}{6\Pi(k)}\delta I^{(3)}(\vk)
 + \frac{M_1}{3 \Pi(k)}  \frac{1}{2 a^2}[(\nabla^2_\vx \varphi - \nabla^2 \varphi)]^{(3)}(\vk) \nonumber\\
 &=  S_1 + \cdots + S_{10}
\end{align}


We compute first for source $S_1$:

\begin{align}
(\T-A(k)) [\Psi_{S_1 \, i,i}^{\typo{(2)}{(3)}}](\vk) &= [\Psi^{(2)}_{i,j} \T \Psi^{(1)}_{j,i}](\vk) 
    = \underset{\vk= \vk_1 + \vk'}{\int} i \frac{k'^{i} k'^j}{k'^{2}} \frac{i}{2}\underset{\vk'= \vk_{23}}{\int} D^{(2)}(\vk_2,\vk_3) \delta_2 \delta_3
      \,\, i^2 \frac{k^j_1 k^i_1}{k_1^2} \T D_+(k_1) \delta_1  \nonumber\\
    & = \frac{1}{6} \ikkk 3 D_+(k_1) A(k_1) D^{(2)}(\vk_2,\vk_3) \frac{(\vk_1 \cdot \vk_{23})^2}{k_1^2 k_{23}^2} \delta_1 \delta_2 \delta_3 ,
\end{align}
then
\begin{align} \label{kPsiS1}
 k_i\Psi_{S_1}^{(3)i}(\vk) &= - \frac{i}{6} \ikkk D^{(3)}_{S_1}(\vk_1,\vk_2,\vk_3) \frac{(\vk_1 \cdot \vk_{23})^2}{k_1^2 k_{23}^2} \delta_1 \delta_2 \delta_3, 
\end{align}
with
\begin{equation}
 D^{(3)}_{S_1} = (\T-A(k))^{-1} \left( 3 A(k_1) D_+(k_1) D^{(2)}(\vk_2,\vk_3) \right) .
\end{equation}

Similarly for the other sources:

\bigskip

\underline{$S_2 = [\Psi^{(1)}_{i,j} \T \Psi^{(2)}_{j,i}](\vk)$:}

\begin{align}
 k_i \Psi_{S_2}^{(3)i}(\vk) &= -\frac{i}{6} \ikkk D^{(3)}_{S_2}(\vk_1,\vk_2,\vk_3) \frac{(\vk_1 \cdot \vk_{23})^2}{k_1^2 k_{23}^2} \delta_1 \delta_2 \delta_3 , 
\end{align}
with
\begin{equation}
 D^{(3)}_{S_2} = (\T-A(k))^{-1} \left( 3 D_+(k_1) \T D^{(2)}(\vk_2,\vk_3) \right). 
\end{equation}

\underline{$S_3 = - A(k)[\Psi^{(2)}_{i,j} \Psi^{(1)}_{j,i}](\vk)$:}

\begin{align}
 k_i \Psi_{S_3}^{(3)i}(\vk) &=  \frac{i}{6} \ikkk D^{(3)}_{S_3}(\vk_1,\vk_2,\vk_3) \frac{(\vk_1 \cdot \vk_{23})^2}{k_1^2 k_{23}^2} \delta_1 \delta_2 \delta_3,
\end{align}
with
\begin{equation}
 D^{(3)}_{S_3} = (\T-A(k))^{-1} \left(  3 A(k) D_+(k_1) D^{(2)}(\vk_2,\vk_3)  \right) .
\end{equation}

\underline{ $S_4 =  - A(k)[\typo{\Psi^{(1)}_{i,j} \Psi^{(2)}_{j,i}}{\Psi^{(1)}_{i,i} \Psi^{(2)}_{j,j}}](\vk)$:}

\begin{align}
  k_i\Psi_{S_4}^{(3)i} (\vk) &=  \frac{i}{6} \ikkk D^{(3)}_{S_4}(\vk_1,\vk_2,\vk_3) \delta_1 \delta_2 \delta_3,
\end{align}
with
\begin{equation}
 D^{(3)}_{S_4} = (\T-A(k))^{-1} \left(  3 A(k) D_+(k_1) D^{(2)}(\vk_2,\vk_3)  \right) .
\end{equation}

\underline{Source $S_5 =  - [\Psi_{i,k}^{(1)}\Psi_{k,j}^{(1)} \T \Psi_{j,i}^{(1)}](\vk)$:}

\begin{align}
  k_i\Psi_{S_5}^{(3)i}(\vk) &= -\frac{i}{6} \ikkk 2\frac{(\vk_1 \cdot \vk_2)(\vk_2 \cdot \vk_3)(\vk_3 \cdot \vk_1)}{k_1^2 k_2^2 k_3^2}
  D^{(3)}_{S_5}(\vk_1,\vk_2,\vk_3)  \delta_1 \delta_2 \delta_3,
\end{align}
with
\begin{equation}
 D^{(3)}_{S_5} = (\T-A(k))^{-1} \left( 3 A(k_1) D_+(k_1)  D_+(k_2)  D_+(k_3)\right) .
\end{equation}

\underline{$S_6 =  \frac{A(k)}{3} [\Psi_{i,k}^{(1)}\Psi_{k,j}^{(1)} \Psi_{j,i}^{(1)}](\vk) $:}

\begin{align} 
  k_i\Psi_{S_6}^{(3)i}(\vk) &= \frac{i}{6}  \ikkk \frac{2}{3} \frac{(\vk_1 \cdot \vk_2)(\vk_2 \cdot \vk_3)(\vk_3 \cdot \vk_1)}{k_1^2 k_2^2 k_3^2}
  D^{(3)}_{S_6}(\vk_1,\vk_2,\vk_3)  \delta_1 \delta_2 \delta_3,
\end{align}
with
\begin{equation}
 D^{(3)}_{S_6} = (\T-A(k))^{-1} \left( 3 A(k) D_+(k_1)  D_+(k_2)  D_+(k_3)\right) .
\end{equation}

\underline{$S_7 = \frac{A(k)}{6} [\Psi_{i,i}^{(1)}\Psi_{j,j}^{(1)}\Psi_{k,k}^{(1)}](\vk)  $:}

%
\begin{align}
  k_i\Psi_{S_7}^{(3)i}(\vk) &= \frac{i}{6} \ikkk \frac{1}{3} D^{(3)}_{S_7}(\vk_1,\vk_2,\vk_3)  \delta_1 \delta_2 \delta_3,
\end{align}
with
\begin{equation}
 D^{(3)}_{S_7} = (\T-A(k))^{-1} \left(  3 A(k) D_+(k_1)  D_+(k_2)  D_+(k_3) \right) .
\end{equation}

\underline{$S_8 = \frac{A(k)}{2} [\Psi_{l,l}^{(1)} \Psi_{i,j}^{(1)} \Psi_{j,i}^{(1)}](\vk)$:}

\begin{align}
  k_i\Psi_{S_8}^{(3)i}(\vk) &= \frac{i}{6} \ikkk \frac{(\vk_2 \cdot \vk_3)^2}{k_2^2 k_3^2}  D^{(3)}_{S_8}(\vk_1,\vk_2,\vk_3)  \delta_1 \delta_2 \delta_3,
\end{align}
with
\begin{equation}
 D^{(3)}_{S_8} = (\T-A(k))^{-1} \left(  3 A(k) D_+(k_1)  D_+(k_2)  D_+(k_3) \right). 
\end{equation}

\underline{$S_9 = \frac{k^2/a^2}{6\Pi(k)}\delta I^{(3)}(\vk)$:}

\begin{align}
  k_i \Psi_{S_9}^{(3)i}(\vk) &= -\frac{i}{6} \ikkk  D^{(3)}_{S_9}(\vk_1,\vk_2,\vk_3)  \delta_1 \delta_2 \delta_3,
\end{align}
with
\begin{equation}
 D^{(3)}_{S_9} = (\T-A(k))^{-1} \left(  \frac{k^2/a^2}{6\Pi(k)} \mathcal{K}_{\delta I}^{(3)} (\vk_1,\vk_2,\vk_3)  D_+(k_1)  D_+(k_2)  D_+(k_3) \right) .
\end{equation}

\underline{$S_{10} = \frac{M_1}{3 \Pi(k)}  \frac{1}{2 a^2}[(\nabla^2_\vx \varphi - \nabla^2 \varphi)]^{(3)}(\vk) $:}

\begin{align}
  k_i \Psi_{S_{10}}^{(3)i}(\vk) &= \frac{i}{6} \ikkk  D^{(3)}_{S_{10}}(\vk_1,\vk_2,\vk_3)  \delta_1 \delta_2 \delta_3,
\end{align}
with
\begin{equation} \label{D2S10}
 D^{(3)}_{S_{10}} = (\T-A(k))^{-1} \left( \frac{M_1(k)}{3 \Pi(k)} \mathcal{K}_{\text{FL}}^{(3)} (\vk_1,\vk_2,\vk_3)  D_+(k_1)  D_+(k_2)  D_+(k_3)\right) .
\end{equation}

\bigskip

Using Eqs.~(\ref{kPsiS1})-(\ref{D2S10}) and rearranging terms we arrive to Eq.~(\ref{LD3order}).

\end{subsection}

\end{section}

 \bibliographystyle{JHEP}  
 \bibliography{refs}        

\begin{thebibliography}{100}
\bibitem{Liaknowled}
{We kindly aknowledge Baojiu Li for providing us with his power spectra
  simulated data}.
\end{thebibliography}

\end{document}